\documentclass[preprint,12pt]{elsarticle}
\usepackage[margin=2.5cm]{geometry} % for preprint

\usepackage{verbatim} % Include the verbatim package for comment
\usepackage[utf8]{inputenc} 

\usepackage{booktabs}
\usepackage{framed,multirow}
\usepackage{hyperref}
\hypersetup{
    colorlinks = true,
    linkcolor = {magenta},
    citecolor = {blue}
}% hyperlinks
\usepackage{url}   % simple URL typesetting

\usepackage{booktabs}% professional-quality tables
\usepackage[mathlines]{lineno}
\usepackage{amsfonts}       % blackboard math symbols
\usepackage{latexsym}
\usepackage{amsmath}
\usepackage{amssymb}
\usepackage{amsfonts}

\usepackage{cleveref}
\usepackage{nicefrac}       % compact symbols for 1/2, etc.
\usepackage{microtype}      % microtypography
\usepackage{lipsum}
\usepackage{adjustbox}
\usepackage{graphicx}
\usepackage{bm}
\usepackage{gensymb}

\usepackage[table,xcdraw]{xcolor}
\definecolor{newcolor}{rgb}{.8,.349,.1}
\usepackage[linesnumbered,ruled,vlined]{algorithm2e}
\usepackage{multirow}
\usepackage{diagbox}
\usepackage{subfig}
\usepackage{tikz}

\usetikzlibrary{external}

\usepackage{pgfplots}
\pgfplotsset{compat=1.7}
\usetikzlibrary{positioning}
\usepackage{etoolbox} %% <- for \pretocmd, \apptocmd and \patchcmd
\colorlet{updated_color}{green!80!red!90!}
\newcommand*\linenomathpatch[1]{%
  \cspreto{#1}{\linenomath}%
  \cspreto{#1*}{\linenomath}%
  \csappto{end#1}{\endlinenomath}%
  \csappto{end#1*}{\endlinenomath}%
}
\newcommand*\linenomathpatchAMS[1]{%
  \cspreto{#1}{\linenomathAMS}%
  \cspreto{#1*}{\linenomathAMS}%
  \csappto{end#1}{\endlinenomath}%
  \csappto{end#1*}{\endlinenomath}%
}

\expandafter\ifx\linenomath\linenomathWithnumbers
  \let\linenomathAMS\linenomathWithnumbers
  \patchcmd\linenomathAMS{\advance\postdisplaypenalty\linenopenalty}{}{}{}
\else
  \let\linenomathAMS\linenomathNonumbers
\fi

\linenomathpatch{equation}
\linenomathpatchAMS{gather}
\linenomathpatchAMS{multline}
\linenomathpatchAMS{align}
\linenomathpatchAMS{alignat}
\linenomathpatchAMS{flalign}

\SetCommentSty{mycommfont}

\crefformat{section}{\S#2#1#3}
\crefformat{subsection}{\S#2#1#3}
\crefformat{subsubsection}{\S#2#1#3}
\crefrangeformat{section}{\S\S#3#1#4 to~#5#2#6}
\crefmultiformat{section}{\S\S#2#1#3}{ and~#2#1#3}{, #2#1#3}{ and~#2#1#3}

\journal{}

\begin{document}

\begin{frontmatter}

\title{A physics-constrained neural solver for incompressible flows without labeled data}

\author[inst1]{\href{https://orcid.org/0009-0002-9465-7208}{\includegraphics[scale=0.08]{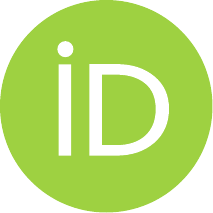}\hspace{1mm}Qifeng Hu
}}
\ead{qih56@pitt.edu}

\author[inst1]{\href{https://orcid.org/0000-0003-1967-7583}{\includegraphics[scale=0.08]{orcid.pdf}\hspace{1mm}Inanc Senocak \corref{cor1}}}
\cortext[cor1]{corresponding author:~senocak@pitt.edu (Inanc Senocak)}

\address[inst1]{Department of Mechanical Engineering and Materials Science, University of Pittsburgh, Pittsburgh, PA 15261, USA}

%\received{xxxx}
%\finalform{xxxx}
%\accepted{xxxx}
%\availableonline{xxxx}
\begin{abstract}
We present a physics-constrained neural solver for incompressible flow that requires neither labeled data nor supervised pretraining. The solver enforces incompressibility and boundary conditions to strict tolerances. We attribute the limited Reynolds numbers attained by existing physics-informed approaches to insufficient enforcement of these constraints, a property on which conventional finite-difference and finite-volume solvers depend. We therefore formulate incompressible flow as a constrained optimization problem within the physics- and equality-constrained artificial neural network (PECANN) framework: the residual of the pressure Poisson equation is minimized subject to the momentum equations, continuity equation, and boundary conditions imposed as equality constraints on the primitive variables, while a conditionally adaptive augmented Lagrangian method enforces each constraint to strict tolerances. The pressure Poisson objective proves critical. Across all benchmarks considered, replacing it with the momentum residual fails to converge to accurate solutions under the same architecture and constraint-enforcement strategy, which we attribute to the global elliptic coupling of pressure and velocity that local momentum-residual minimization lacks. For advection-dominated flows, we introduce an adaptive vanishing entropy viscosity that stabilizes training while leaving the converged solution unchanged. The solver predicts lid-driven cavity flow up to $Re = 7,500$ with centerline velocity profiles matching benchmark data, and is further validated on flow over a backward-facing step at $Re = 800$, three-dimensional unsteady Beltrami flow, and steady and unsteady flow past a circular cylinder under an inflow–outflow configuration, for which an ablation study identifies admissible outlet boundary conditions for velocity and pressure. Starting from a randomly initialized network and without external perturbations, the solver captures the spontaneous onset of periodic vortex shedding in unsteady cylinder flow.
\end{abstract}

%% Keywords %%

\end{frontmatter}
%\linenumbers

%%%% Introduction %%% 
\section{Introduction}\label{sec:intro}
Multi‑layer feed‑forward neural networks were applied to the solution of differential equations \citep{dissanayake1994neural, MEADE199419, van1995neural} a few years after they were established as universal approximators \citep{hornik1989multilayer}. With rapid advances in the field of machine learning—driven by improvements in both hardware and software platforms—this original idea has been revitalized in the form of physics‑informed neural networks (PINNs) \citep{raissi2019pinn}, which have since seen widespread adoption across a broad range of scientific and engineering disciplines \citep{Karniadakis2021}.

PINNs and their various extensions have been applied to several classes of PDEs and, with limited success, to the incompressible Navier–Stokes equations. However, to date, there has been no successful demonstration of solving incompressible flow equations using neural networks in a purely unsupervised fashion — without auxiliary (labeled) data, pre-training on labeled flow solutions, or analytical reference solutions — analogous to conventional finite‑difference or finite‑volume solvers. We attribute this gap to the absence of flow‑solution algorithms specifically tailored for neural‑network implementations. To address this limitation, we introduce a neural‑network‑based method for solving the incompressible flow equations without any such dependencies.

The simulation of high‑Reynolds-number incompressible flows over complex geometries has long been, and continues to be, a central focus of computational fluid dynamics (CFD). The success of incompressible flow solvers depends critically on strict enforcement of a divergence-free velocity field through the solution of a Poisson equation for the pressure or a pressure-correction variable. Although a streamfunction–vorticity formulation can bypass the need to calculate the pressure field explicitly, such formulations are practically limited to two-dimensional flows. Consequently, most three-dimensional solvers employ the primitive-variable form of the Navier–Stokes equations. Conventional incompressible flow solvers — including projection methods \citep{chorin1968projection, Temam1969, Kim1985}, pressure-correction methods \citep{patankar1972, issa1986piso}, and spectral element formulations \citep{karniadakis_incompressible_1991} — share this common algorithmic feature, and a comprehensive analysis of their implementations and implications for numerical accuracy can be found in \citet{guermond_review_projection}. 

PINNs have emerged as an alternative to conventional mesh-based numerical methods for solving PDEs. PINNs have been applied to simulate incompressible flows at low to modest Reynolds numbers. Advection-dominated regimes characterized by high Reynolds number remain a major challenge. Here, we focus on studies that adopt PINNs with velocity–pressure formulation of the Navier-Stokes equations and exclude works that use streamfunction–vorticity formulation. 

\citet{jiang_2023} applied PINNs to the lid-driven cavity benchmark at $Re \in \{100, 1000, 10000\}$. Notable discrepancies from reference velocity profiles even at $Re=100$ prompted the authors to augment the standard PINN approach with finite differences (FD-PINN). Although FD-PINN achieved excellent agreement at $Re=100$, predictions deteriorated significantly at $Re=1,000$, while the original PINN approach failed altogether. Predictions were improved for both approaches when random labeled data from a reference simulation were added to their training, while the original PINN approach showed deficiencies at higher Reynolds number cases.

\citet{jin_nsfnets_2021} introduced Navier–Stokes flow nets (NSFnets), exploring both velocity–pressure (VP) and vorticity–velocity (VV) formulations. While NSFnets can reproduce solutions for low‑Reynolds-number flows, they require initial and boundary conditions obtained from analytical solutions or high-fidelity simulations, and also benefit from transfer learning in which a network trained at low Reynolds number is repurposed for higher Reynolds number. For turbulent channel flow, DNS data were required to supply initial and boundary conditions, and the VV formulation produced inconsistent results in this regime.

\citet{roy_2024_cavity_flow} tested the velocity-pressure and streamfunction-pressure formulations of the incompressible flow equations using PINNs, enforcing all boundary conditions exactly through a hard-mask. For the lid-driven cavity benchmark, the velocity-pressure formulation failed to produce acceptable results beyond $Re=400$, whereas the streamfunction-pressure formulation produced good results for Reynolds numbers as high as $Re=5000$. This stark difference in performance points to the importance of enforcing a divergence-free condition — a condition the streamfunction-pressure formulation satisfies by design, albeit only in two dimensions. \citet{roy_2024_cavity_flow} also showed that the velocity-pressure formulation can produce good results at higher Reynolds numbers if labeled data from a high-fidelity simulation are introduced inside the domain. 

\citet{SHUKLA2024117290} compared PINNs and physics-informed Kolmogorov-Arnold networks with Chebyshev polynomials (cPIKANs) on the lid-driven cavity benchmark. While both approaches produced solutions in agreement with reference data at $Re=400$, training became unstable at $Re=2,000$, prompting the authors to introduce an entropy viscosity for stabilization. Validation was limited to streamline comparisons against reference data. The PINN method captured the overall flow structure but showed visible discrepancies in the corner vortices, while cPIKAN produced an entirely different circulation pattern in the cavity, suggesting that increasing the expressive capacity of a network is not sufficient to produce accurate predictions for high Reynolds number flows.
 
\citet{wang2024piratenet} introduced the physics-informed residual adaptive networks (PirateNets) and simulated lid-driven cavity flow up to $Re = 3200$
 using a primitive-variable formulation, reporting excellent agreement with benchmark data. To address training instability at high Reynolds numbers, the authors adopted a curriculum strategy in which the Reynolds number was gradually increased ($Re=100\rightarrow400\rightarrow1000\rightarrow1600\rightarrow3200$), and employed a deep network with 18 layers and 256 channels --- both of which significantly increased the overall computational cost.
 
PirateNets --- combined with causality training \citep{Wang2024causality} and modern optimization techniques \citep{vyas2025soap} --- were applied to simulate three-dimensional turbulent channel flow at $Re_\tau = 395$ \citep{wang2025piratenetturbulence}. A snapshot from a reference direct numerical simulation (DNS) was used as the initial condition, and periodic boundary conditions in the streamwise and spanwise directions were imposed as hard constraints. Turbulence was maintained over a finite time horizon. While the mean velocity profiles showed excellent agreement with DNS, the mean Reynolds stress components exhibited noticeable discrepancies, particularly near their peak values, and the energy spectra showed systematically lower energy across the resolved wavenumber range, suggesting excessive dissipation in the learned solutions. Notably, momentum and continuity residuals on the order of $10^{-3}$ likely contributed to these discrepancies in second-order statistics and energy spectra.

To the best of our knowledge, no existing neural-network based approach can reliably solve the incompressible flow equations for general inflow-outflow configurations at high Reynolds numbers without labeled data or supervised pretraining. We believe this limitation has a clear root cause: unlike conventional CFD solvers, which rigorously enforce the divergence-free constraint and boundary conditions, existing PINN approaches lack a principled mechanism for constraint enforcement. Without it, the network has no guarantee of producing physically consistent solutions.

To address this gap directly, we adopt the physics- and equality-constrained artificial neural network (PECANN) framework \citep{PECANN_2022} with the conditionally adaptive augmented Lagrangian method (CA-ALM) \citep{hu2026caalm}, which enforces heterogeneous constraints to strict tolerances in a systematic and adaptive way. Within this rigorous constrained optimization framework, we introduce an objective function tailored for incompressible flows: the residual of the pressure Poisson equation is minimized subject to the momentum and continuity equations and boundary conditions on the primitive variables. This formulation ensures that both the pressure and velocity fields are physically consistent by construction.

We assess the performance of the proposed method on several benchmark problems: 2D lid-driven cavity flow up to $Re\leq 7500$, 3D Beltrami flow, and steady and unsteady flow past a circular cylinder with general inflow–outflow boundary conditions. No labeled data are used and all networks are trained from a random distribution.

%%%% Methodology %%% 
\section{Methodology}\label{sec:method}
Constant density incompressible flows in the spatio-temporal domain $\mathcal{U} = \Omega \times [0, T]$ are governed by the Navier-Stokes equations together with the continuity equation in divergence-free form:
\begin{subequations}
\begin{align}
    \bm{\mathcal{F}}(\bm{x},t) & = \frac{\partial \bm{u}}{\partial t} + (\bm{u} \cdot \nabla) \bm{u} + \nabla p - \nu \nabla^2 \bm{u} =0 & \quad \text{in} \quad \mathcal{U}, \label{eq:uns_incompress_mom}\\
    \mathcal{M}(\bm{x},t) & = \nabla \cdot \bm{u} = 0 & \quad \text{in} \quad \mathcal{U}. \label{eq:uns_incompress_mass}
\end{align}
\label{eq:uns_incompress_eq}
\end{subequations}
Here, $\bm{x}=[x,y,z]^\top$ denotes the spatial coordinate vector, $t$ the time, $\bm{u}(\bm{x},t) = [u, v, w]^\top$ and $p(\bm{x},t)$ the velocity and pressure fields defined over $\mathcal{U}$. The kinematic viscosity is denoted by $\nu$,  and the Reynolds number is defined as $Re= U L /\nu$, where $U$ and $L$ are characteristic velocity and length scales, respectively. Unit density is assumed throughout.

The operator $\bm{\mathcal{F}} = [\mathcal{F}_u, \mathcal{F}_v, \mathcal{F}_w]^\top$ in Eq.~\eqref{eq:uns_incompress_mom} represents the three components of the momentum conservation equations. The operator $\mathcal{M}$ in Eq.~\eqref{eq:uns_incompress_mass} enforces the divergence-free condition.

These governing equations are supplemented with boundary conditions on $\partial \Omega\times[0,T]$ and initial conditions on $\Omega \times \{t=0\}$, typically expressed in terms of Dirichlet or Neumann operators:
\begin{subequations}
\begin{align}
    \mathcal{D}(\bm{x}, t) & = \phi(\bm{x}, t) - g(\bm{x}, t) = 0 & \quad \text{on} \quad \partial_D \mathcal{U}, \\
    \mathcal{N}(\bm{x}, t) & = \frac{\partial \phi(\bm{x}, t)}{\partial \bm{n}}  - h(\bm{x}, t) = 0 & \quad \text{on} \quad \partial_N \mathcal{U},
\end{align}
\end{subequations}
where $\phi$ represents any components of velocity or pressure field, $\bm{n}$ is the outward unit normal vector, and $\partial_D \mathcal{U}$ and $\partial_N \mathcal{U}$ denote Dirichlet and Neumann portions of $\partial \mathcal{U}$, respectively.

In CFD practice, Dirichlet conditions prescribe the dependent variable directly through $g(\bm{x},t)$, while Neumann conditions impose its normal derivative or flux $h(\bm{x},t)$. For fluid flows, typical examples include no-slip walls, free-slip walls, inlet condition  with a given velocity profile. We note that specific boundary conditions on the velocity and pressure field are discussed in detail in the Section \ref{sec:results}.

If time evolution of a flow field from a given state is of interest, initial conditions must be specified, typically expressed using Dirichlet operators:  
\begin{equation}
    \mathcal{I}_\phi(\bm{x}, t) = \phi(\bm{x}, t) - \phi_0(\bm{x}) = 0 
    \quad \text{in} \quad \Omega \times \{t=0\},
\end{equation}
where $\phi_0(\bm{x})$ denotes the prescribed initial distribution of the field $\phi$ at $t=0$.

\subsection{Proposed method: pressure-Poisson-based objective function}
Strict enforcement of the continuity equation and boundary conditions is essential for accurate incompressible flow simulations. Unlike compressible flows—where pressure is a thermodynamic quantity linked to density and temperature via an equation of state—pressure in incompressible flows is purely hydrodynamic with no governing equation of its own. Instead, a Poisson equation for pressure is derived by taking the divergence of the momentum equations and enforcing a solenoidal velocity field:
\begin{equation}
\mathcal{P}(\bm{x}) = \nabla^2 p + \big(\nabla \bm{u}\big) : \big(\nabla \bm{u}\big)^{\top} = 0 
\quad \text{in} \quad \mathcal{U}.
\label{eq:psn_p}
\end{equation}

Since Eq.~\eqref{eq:psn_p} is strictly conditioned on the satisfaction of the governing equations~\eqref{eq:uns_incompress_eq} due to solenoidal velocity field, we propose equality-constrained optimization problem for learning incompressible flows by a single neural network parameterized by $\theta$ to approximate the true velocity and pressure fields without splitting the momentum equations:
\begin{equation}
    [\hat{\bm{u}}, \hat{p}]^\top(\bm{x},t;\theta) \approx [\bm{u}, p]^\top.
\end{equation}
In this formulation, the mean square residual of the pressure Poisson equation is adopted as the objective function $\mathcal{J}$. This objective is constrained by the residuals of the momentum equations in each direction $\mathcal{C}_{F,k}$ ($k \in \{u, v, w\}$), along with the divergence-free condition, $\mathcal{C}_M$, Dirichlet $\mathcal{C}_{D_j}$, and Neumann $\mathcal{C}_{N_j}$ type boundary conditions, and initial conditions $\mathcal{C}_{I_\phi}$. Following \citep{hu2026caalm}, a mean-squared-residual metric is adopted as a constraint aggregation technique to avoid pointwise enforcement of the constraints.

The constrained optimization problem is expressed as follows:
\begin{equation}
\begin{aligned}
    \min_{\theta} \quad & \mathcal{J}(\theta; \bm{x}, t) = \frac{1}{N_{\mathcal{U}}} \sum_{i=1}^{N_{\mathcal{U}}} \|\mathcal{P}(\theta; \bm{x}_i, t_i)\|_2^2 && ~\text{in} \quad \mathcal{U}, \\
    \text{subject to} \quad & \mathcal{C}_{F,k}(\theta; \bm{x}, t) = \frac{1}{N_{\mathcal{U}}} \sum_{i=1}^{N_{\mathcal{U}}} \|\mathcal{F}_k(\theta; \bm{x}_i, t_i)\|_2^2 = 0 && ~\text{in} \quad \mathcal{U}, \\
        & \mathcal{C}_M(\theta; \bm{x},t) = \frac{1}{N_{\mathcal{U}}} \sum_{i=1}^{N_{\mathcal{U}}} \|\mathcal{M}(\theta; \bm{x}_i, t_i)\|_2^2 = 0 && ~\text{in} \quad \mathcal{U}, \\
        & \mathcal{C}_{D_j}(\theta; \bm{x}, t) = \frac{1}{N_{D_j}} \sum_{i=1}^{N_{D_j}} \| \mathcal{D}(\theta; \bm{x}_i, t_i) \|_2^2 = 0 && ~\text{on} \quad D_j \subseteq \partial_D \mathcal{U}, \\
        & \mathcal{C}_{N_j}(\theta; \bm{x}, t) = \frac{1}{N_{N_j}} \sum_{i=1}^{N_{N_j}} \| \mathcal{N}(\theta; \bm{x}_i, t_i) \|_2^2 = 0 && ~\text{on} \quad N_j \subseteq \partial_N \mathcal{U}, \\
        & \mathcal{C}_{I_\phi}(\theta; \bm{x}, t) = \frac{1}{N_{I}} \sum_{i=1}^{N_{I}} \| \mathcal{I_\phi}(\theta; \bm{x}_i, t_i) \|_2^2 = 0 && ~\text{on} \quad \Gamma.
    \label{eq:proposed_constrained_flow_problem}
\end{aligned}
\end{equation}
Here, $j \in \{u,v,w,p\}$ enumerates the primitive variables at the boundary. 

This equality-constrained formulation provides a physically meaningful performance metric: the objective directly measures how well the pressure and velocity fields satisfy the pressure Poisson equation, while the constraints enforce adherence to momentum equations in each direction, divergence-free condition, and boundary conditions within strict tolerances.

\begin{figure}[!htb]
    \centering
    \includegraphics[width=1.0\linewidth]{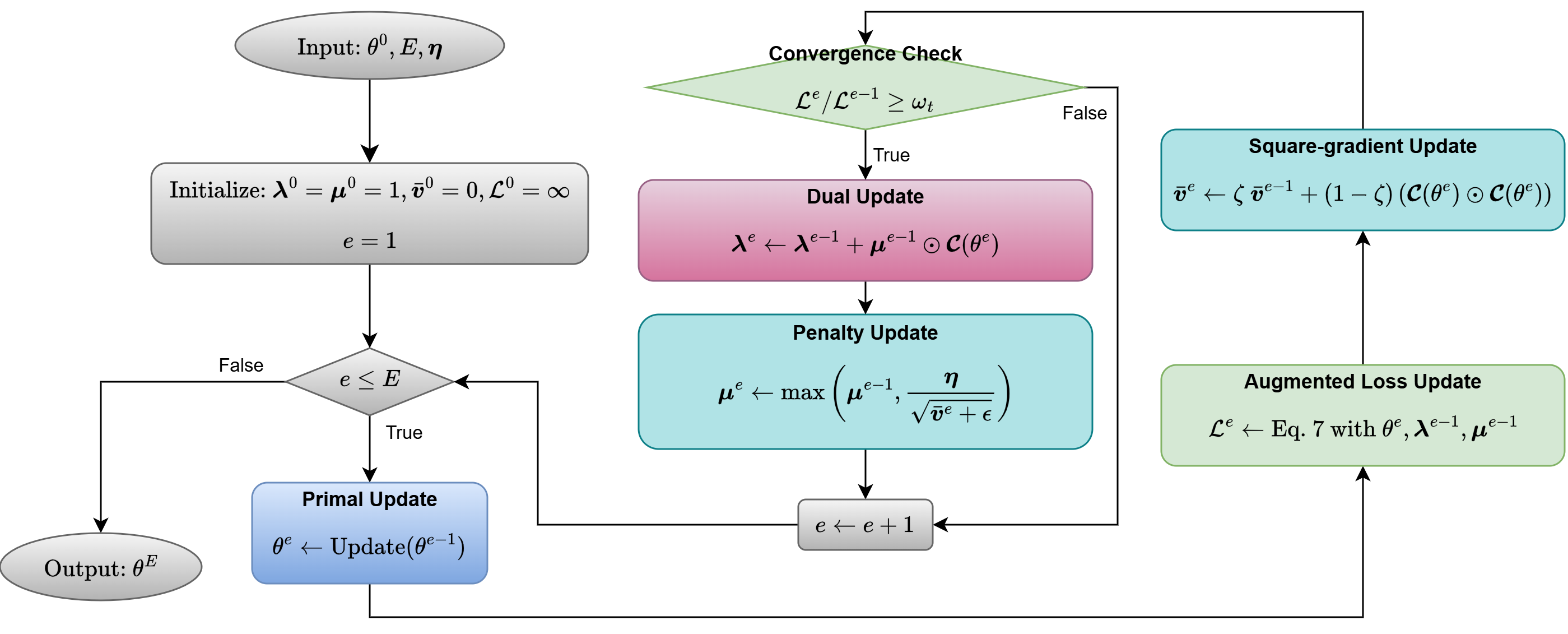}
    \caption{Flowchart of the Conditionally Adaptive Augmented Lagrangian method (CA-ALM) \cite{hu2026caalm}.}
    \label{fig:caalm_alg}
\end{figure}

%  \cite{hestenes1969multiplier, powell1969method}
Eq. \ref{eq:proposed_constrained_flow_problem} is a challenging optimization problem with heterogeneous constraints. We employ the conditionally adaptive augmented Lagrangian method (CA-ALM) \cite{hu2026caalm}, which assigns independent Lagrange multipliers and penalty parameters to each constraint and updates them separately. CA-ALM recasts the constraint optimization problem into the following unconstrained form:
\begin{align}
    \max_{\bm{\lambda}} \; \min_{\theta} \; 
    \mathcal{L}(\theta,\bm{\lambda};\bm{\mu}) 
    &= \mathcal{J}(\theta) 
    + \bm{\lambda}^\top \bm{\mathcal{C}}(\theta) 
    + \tfrac{1}{2}\, \bm{\mu}^\top \big[\bm{\mathcal{C}}(\theta) \odot \bm{\mathcal{C}}(\theta)\big],
    \label{eq:proposed_max_min}
\end{align}
where $\bm{\mathcal{C}}(\theta)$ collects all constraints into a vector, $\odot$ denotes the elementwise (Hadamard) product, $\bm{\lambda}$ is the vector of Lagrange multipliers, and $\bm{\mu}$ contains the positive penalty parameters, with $\mu_i$ associated with constraint $\mathcal{C}_i$. The complete training procedure for updating $\theta, \bm{\lambda}, \bm{\mu}$ are presented in vectorized form in Figure~\ref{fig:caalm_alg} and the algorithm for CA-ALM in pseudo-code is available in \citet{hu2026caalm}.

In Eq.~\eqref{eq:proposed_constrained_flow_problem}, multiple constraints are present, each associated with a penalty scaling factor $\eta_i \in \bm{\eta}$ in Fig.~\ref{fig:caalm_alg}, typically chosen from $\{10^{-2}, 10^{-1}, 1\}$. These factors are used to reflect the relative priority of constraints during optimization. In particular, the scaling factors for the PDE constraints should be chosen smaller than those for boundary and initial conditions. This prioritization ensures boundary and initial conditions are satisfied early in the optimization before reducing the residuals of the governing PDEs. Consequently, we set $\eta_i=1$ for the boundary and initial conditions, as well as for the divergence-free velocity constraint $\mathcal{C}_M$, while assigning smaller values (e.g., $\eta_i=0.1$) to the momentum equations.

\subsection{Baseline (Na{\"i}ve) formulation: momentum-equation-based objective function}
To demonstrate the merits of the proposed formulation \eqref{eq:proposed_constrained_flow_problem}, we introduce a baseline equality-constrained problem that omits the pressure Poisson equation from the objective. Here, the objective $\mathcal{J}_b$ minimizes the cumulative momentum losses $\mathcal{C}_{F,k}$, while all remaining constraints in Eq.~\eqref{eq:proposed_constrained_flow_problem} are enforced via CA-ALM:
\begin{equation}
\begin{aligned}
    \min_{\theta} \quad & \mathcal{J}_b(\theta; \bm{x}, t) = \sum_k \mathcal{C}_{F,k}(\theta; \bm{x}, t) && ~\text{in} \quad \mathcal{U}, \\
    \text{subject to} \quad & \bm{\mathcal{C}}(\theta) = [\mathcal{C}_M(\theta; \bm{x},t), \,\mathcal{C}_{D_j}(\theta; \bm{x}, t), \,\mathcal{C}_{N_j}(\theta; \bm{x}, t), \,\mathcal{C}_{I_\phi}(\theta; \bm{x}, 0)]^\top = \textbf{0}.
    \label{eq:baseline_constrained_flow_problem}
\end{aligned}
\end{equation}
Minimizing the coupled momentum losses in the objective requires controlled growth of the Lagrange multipliers and penalty parameters. The baseline formulation is less tolerant of large penalties — to maintain stable, accurate predictions, its scaling factors are set to $0.01$ times those of the proposed formulation in this study. %Specific values of the hyperparameters used in CA-ALM are defined in Section \ref{sec:results}.

\subsection{Adaptive vanishing entropy viscosity}
High-order numerical schemes are prone to numerical instability when applied to advection-dominated flows characterized by large Reynolds numbers. A common remedy is to augment the physical viscosity $\nu$ with an artificial viscosity term $\nu_a$ \citep{guermond2011evm, wang2019evm}. 
%such that the effective viscosity becomes
%\begin{equation}
%\nu = \frac{1}{Re} + \nu_a,
%\end{equation}
%thereby relaxing the near-zero diffusion regime \citep{guermond2011evm, wang2019evm}. 
In the context of PINNs, artificial viscosity was found to stabilize the training process \citep{he_artificial_2023}. In a comprehensive comparison of physics-informed learning approaches, \citet{SHUKLA2024117290} treated the entropy residual $r(\bm{x},t)$ as an additional network output to derive a non-vanishing entropy viscosity similar to that of \citet{wang2019evm}, with the key difference that their formulation retained a non-zero entropy viscosity near wall boundaries to improve predictions   

In contrast, we introduce a vanishing entropy viscosity solely to stabilize the early stages of training, with no influence on the final solution. We compute the entropy residual directly from the predicted velocity and pressure field as
\begin{equation}
r(\bm{x},t) = \left(\bm{u}(\bm{x},t) - \bm{u}_m\right) \cdot \bm{\mathcal{F}}(\bm{x},t),
\end{equation}
where $\bm{u}_m$ represents a reference velocity vector.
The infinity norm $\|r\|_\infty$ is used to define a global, rather than pointwise, entropy viscosity. We further make this entropy viscosity adaptive by adjusting it dynamically based on optimization feedback. 

Following the same procedure as CA-ALM, (Fig.~\ref{fig:caalm_alg}), we maintain an exponential moving average of the squared entropy norm $(||r||_\infty)^2$ and use its unbiased root to update the entropy viscosity $\nu_a^a$ at epoch $e$:
\begin{equation}
    \begin{aligned}
        \kappa^{(e)} & \leftarrow \zeta_a \kappa^{(e-1)} + (1-\zeta_a) (||r||_\infty)^2, \\
        \nu_a^{a(e)} & \leftarrow \min(\nu_a^{a(e-1)}, \frac{\sqrt{\kappa^e}}{1-\zeta_a} \frac{L^2}{U^2}||r||_\infty ), \text{if } \nu_a^{a(e-1)} > \nu_a^{a(f)}, & \text{otherwise } 0,
    \end{aligned}
    \label{eq:adapt_av_update}
\end{equation}
where $U$ and $L$ are characteristic velocity and length scales used in defining the Reynolds number, respectively. $\zeta_a=0.99$ is the smoothing coefficient, the same value as one in CA-ALM algorithm. This entropy viscosity is applied exclusively to high-Reynolds-number cases, with initial value $\nu_a^{a(0)} = \tfrac{1}{200}$. The initial value $\kappa^{(0)} = 1$ is chosen to prevent an abrupt drop from $\nu_a^{a(0)}$, while the bias-correction factor $1/(1-\zeta^e)$ prevents $\nu_a^a$ from being artificially inflated in early epochs. \textcolor{black}{Once $\nu_a^a$ falls below the $Re$-dependent cutoff $\nu_a^{a(f)}$, taken as $\tfrac{\nu}{4}$ by default, the remaining artificial dissipation is small enough that removing it does not destabilize the optimization, and $\nu_a^{a}$ is set identically to zero for the remainder of training, and the converged solution reflects the true Reynolds number of the flow. 
This formulation is referred to hereafter as the adaptive vanishing viscosity. \ref{sec:app_d} presents an ablation study of this formulation.}

To assess the merits of the proposed adaptive vanishing viscosity, we also consider a global version of the non-vanishing point-wise formulation of entropy viscosity from \citet{SHUKLA2024117290}, which defines a linear relation between the artificial viscosity and the infinity norm of the entropy residual:
\begin{equation}
\nu_a^l = \min \left(\beta \frac{1}{Re}, \alpha \frac{L^2}{U^2} ||r||_\infty \right), \label{eq:global_artificial_visc_a}
\end{equation}
where $\beta$ and $\alpha$ are prescribed constants controlling the magnitude of the artificial viscosity. When adopted within the PECANN framework, predictions were found to be highly sensitive to the choice of $\beta$ and $\alpha$, motivating the adaptive vanishing formulation introduced earlier. This formulation is referred to hereafter as the linear entropy viscosity. 

\subsection{Fourier feature mappings}
A Fourier feature mapping, $\gamma$, projects input coordinates $\mathbf{x} \in \mathbb{R}^d$ into a higher-dimensional space via sinusoidal transformations, thereby enhancing the expressive capacity of MLP networks \cite{tancik_fourier_2020, wang_eigenvector_2021}. This technique is theoretically grounded in Bochner’s theorem and can be interpreted as a Fourier approximation of a stationary kernel function \cite{tancik_fourier_2020}, enabling the network to capture high-frequency components in the input space more effectively than MLPs.

A single Fourier feature mapping is defined as:
\begin{equation}
\gamma(\mathbf{x}) =
\begin{bmatrix}
\cos(2\pi \mathbf{B} \mathbf{x}) \\
\sin(2\pi \mathbf{B} \mathbf{x})
\end{bmatrix}
\end{equation}
where $\mathbf{B} \in \mathbb{R}^{m \times d}$ consists of entries sampled from a Gaussian distribution $\mathcal{N}(0, \sigma)$, which remain fixed during training. 

In contrast to regions of large velocity variation, the free-stream region approaches a uniform velocity that varies trivially with the input coordinates. The randomly sampled Fourier frequencies may not adequately cover the near-linear behavior of such regions. We therefore concatenate the original input $\mathbf{x}$ with the Fourier features to retain direct access to the input coordinates and ensure the representation of these nearly zero-frequency components. With this modification, the first hidden layer is formed as:
\begin{equation}
\mathbf{H}_1 = \begin{bmatrix} \gamma(\mathbf{x}) \\ \mathbf{x} \end{bmatrix},
\label{eq:modified_1st_hidden}
\end{equation}
and the following layer-wise operations take the following forms:
\begin{equation}
\begin{aligned}
\mathbf{H}_\ell &= \psi(\mathbf{W}_\ell \cdot \mathbf{H}_{\ell-1} + \mathbf{b}_\ell), \quad \ell = 2, \ldots, H, \\
f_\theta(\mathbf{x}) &= \mathbf{W}_{L+1} \cdot \mathbf{H}_L + \mathbf{b}_{L+1},
\end{aligned}
\label{eq:fourier_net}
\end{equation}
where $\theta = \{\mathbf{W}_\ell, \mathbf{b}_\ell, \ldots\}$ and $\psi$ represents the activation function. Following the empirical demonstration of PECANN \citep{hu2026caalm} for PDEs with highly oscillatory solution, the standard deviation of the Gaussian distribution $\sigma$ is set to 1.0 in the present work as well. Additionally, the conventional MLP network architecture corresponds to replacing \eqref{eq:modified_1st_hidden} with a standard affine transformation:
\begin{equation}
    \mathbf{H}_1 = \mathbf{W}_1 \mathbf{x} + \mathbf{b}_1.
\end{equation}

%forward solution 
\section{Application to internal and external flow scenarios}\label{sec:results}
We apply our proposed method to the following benchmark incompressible flow problems: 2D lid-driven cavity (LDC) flow up to a Reynolds number of 7500, 3D time dependent Beltrami flow, 2D steady and unsteady flow over a cylinder at Reynolds number of 40 and 100, respectively. 

Given a vector of predictions $\boldsymbol{\hat{u}} \in \mathbf{R}^n$ and an $n$-dimensional vector of exact values $\boldsymbol{u} \in \mathbf{R}^n$, we define the relative Euclidean or $\ell^2$ norm to assess the accuracy of our predictions. 
\begin{equation}
  \mathcal{E}_r(\hat{\boldsymbol{u}}, \boldsymbol{u})
  = \frac{\|\hat{\boldsymbol{u}} - \boldsymbol{u}\|_2}
         {\|\boldsymbol{u}\|_2},
  \label{eq:relative_L2_error}
\end{equation}
where $\|\cdot \|_2$ denotes the Euclidean norm.

Unless otherwise stated, we use artificial neural networks with hyperbolic tangent activation functions and Xavier initialization scheme \cite{glorot2010understanding}. L-BFGS optimizer available in the Pytorch package \cite{paszke2019pytorch} is employed with \emph{strong Wolfe} line search function. For evaluations involving statistical performance, such as the relative $\ell^2$ norm, the mean prediction and its standard deviation are based on five independent trials.

\subsection{Two-dimensional steady lid-driven cavity flow}
\label{sec:lid_driven_cavity_flow}
The 2D LDC flow is a classical benchmark for incompressible flow solvers in advection-dominated regimes. Its closed domain enables rigorous mass conservation verification, with well-established reference data available from steady numerical simulations up to $Re=10,000$ in \citet{ghia_1982} and $Re=21,000$ in \citet{erturk_numerical_2005}. Hydrodynamic stability analyses on the LDC flow identified a Hopf bifurcation at $Re \approx 8,000$ \citep{auteri2002_LDC_bifurcation, BRUNEAU2006_LDC}, beyond which the flow becomes time-periodic. We therefore restrict our simulations to the steady regime, up to $Re=7,500$.

While finite-difference and finite-volume methods match these benchmarks with high fidelity, many alternative numerical approaches that perform well at $Re=100$
and $Re=400$ exhibit significant discrepancies for $Re \gtrsim 1000$, reflecting the difficulty of enforcing mass conservation in advection-dominated flow regimes. Realizing the challenge of high Reynolds number regime, we simulate the steady LDC problem at $Re \in \{2500, 5000, 7500\}$.

The problem is defined in a closed square cavity $\Omega = \{(x,y) | 0 \leq x, y \leq 1\}$ with the the top lid moving with a prescribed velocity in the $x$-direction, thus involving only Dirichlet boundary conditions.
As the Reynolds number increases, the flow develops distinct corner vortices, and the predicted velocity profiles are commonly validated against reference data \cite{ghia_1982, erturk_numerical_2005}.

We simulate the $Re = 2500$ case to perform a comprehensive ablation study comparing our pressure-Poisson-based and the baseline (na{\"i}ve) formulations of the objective function in the PECANN framework. Two additional design choices are considered: the network architecture (Fourier-feature network vs. MLP network) and the entropy viscosity strategy (linear $\nu_a^l$ vs. adaptive vanishing $\nu_a^a$). Validation results for $Re=5,000$ and $Re=7,500$ are presented in~\ref{sec:app_a} and~\ref{sec:app_b}, respectively.

Unlike finite-difference methods on a structured Cartesian mesh, the velocity discontinuities at the intersections of the moving lid with the sidewalls generate large, localized residuals that hinder optimizer convergence. To alleviate this issue, a smooth lid velocity profile is adopted following \citet{wang2024piratenet}:
\begin{equation}
u(x, 1) = 1 - \frac{\cosh[50(x - 0.5)]}{\cosh(25)}, \quad v(x, 1) = 0.
\end{equation}
Additionally, since only the spatial derivatives of the pressure field appear in the governing equations, the pressure field is anchored to an arbitrary reference value at an arbitrary point within the domain ($p(0.5, 0.5) = 0$) to prevent its unbounded growth. This single-point constraint $\mathcal{C}_p$ is enforced with its own penalty scaling factor of $\eta_p = 0.1$. The constrained optimization problem for the pressure-Poisson-based formulation is then written as 
\begin{equation}
    \min_\theta \mathcal{J}(\theta; \bm{x}), \quad \text{subject to} \quad \mathcal{\bm{C}}(\theta; \bm{x}) = [\mathcal{C}_{F,u},\, \mathcal{C}_{F,v},\, \mathcal{C}_M,\, \mathcal{C}_{B,u},\, \mathcal{C}_{B,v},\, \mathcal{C}_{p}]^\top = \textbf{0}.
\end{equation}
For comparison, the baseline formulation is expressed as
\begin{equation}
    \min_\theta \mathcal{J}_b(\theta; \bm{x})=\mathcal{C}_{F,u}+ \mathcal{C}_{F,v}, \quad \text{subject to} \quad \mathcal{\bm{C}}(\theta; \bm{x}) = [\mathcal{C}_M,\, \mathcal{C}_{B,u},\, \mathcal{C}_{B,v},\, \mathcal{C}_{p}]^\top = \textbf{0},
\end{equation}
which omits the pressure Poisson equation and directly minimizes the sum of the residuals of the momentum equations in each direction constrained by the boundary conditions on the velocity field, divergence-free condition, and anchoring of the pressure field at a single point.

%\subsubsection{Ablation study}
We employ neural networks with four hidden layers, each containing 80 neurons (for the Fourier-feature network, the first hidden layer corresponds to the Fourier features mapping).
Each model is trained for a total of 100{,}000 epochs. A total of 20{,}000 residual points are randomly sampled within the domain, and 256 points are sampled along each boundary.
For the linear entropy viscosity, $\nu_a^l$, (Eq. \ref{eq:global_artificial_visc_a}), the characteristic length and velocity are set to $L = 1$ and $U = 1$, respectively. The reference velocity is defined as $\bm{u}_m = (0.5, 0.5)$, and the prescribed constants $\beta = 5$ and $\alpha = 0.03$ as adopted in \cite{SHUKLA2024117290}.
\begin{figure}[!htb]
    \centering
    \includegraphics[width=0.9\linewidth]{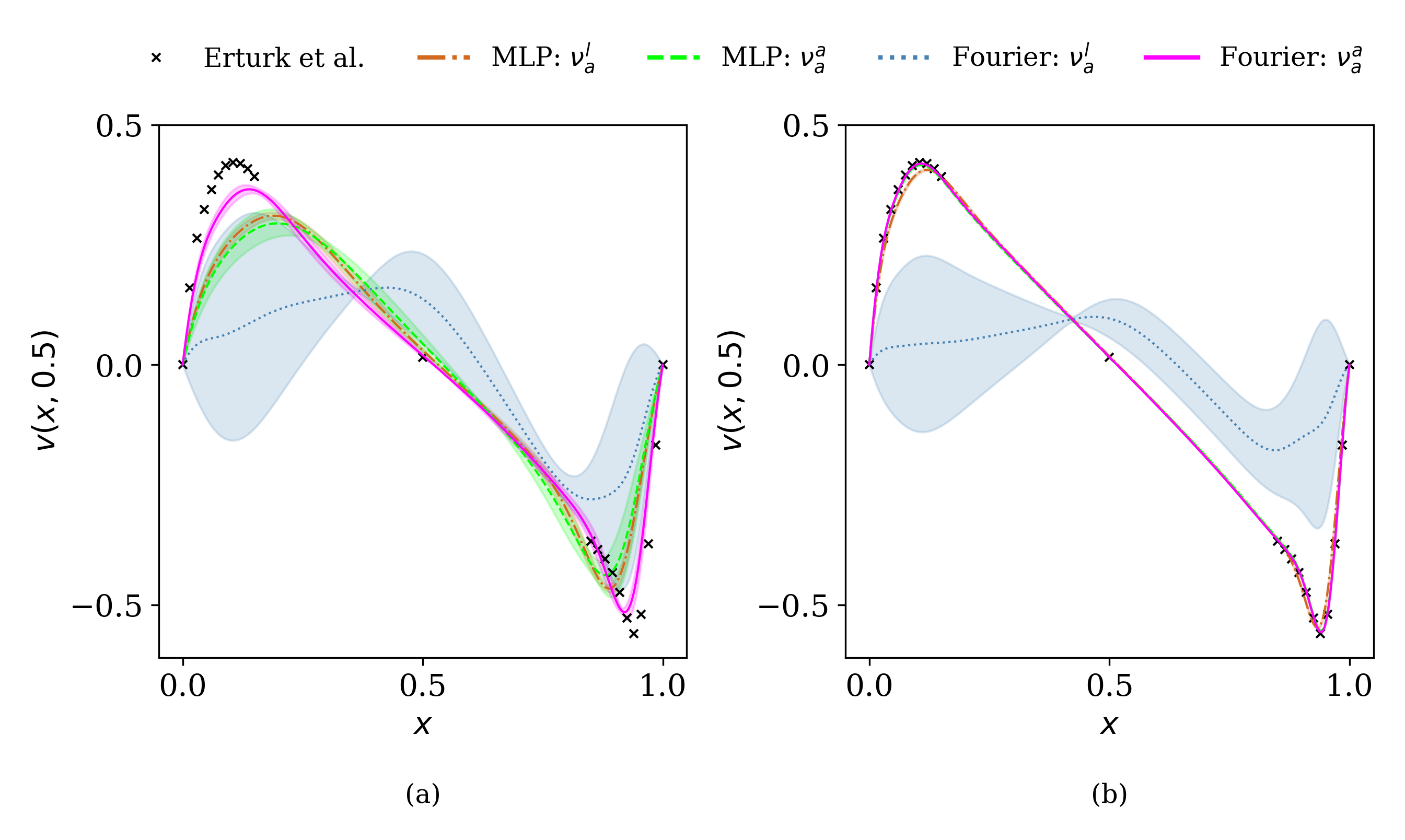}
    \caption{Lid-driven cavity flow at $Re = 2500$: mean and standard deviation (Std. band) of the predicted $v$-velocity along $y = 0.5$ for (a) the baseline formulation (Eq.~\ref{eq:baseline_constrained_flow_problem}) and (b) the proposed pressure-Poisson-based  formulation (Eq.~\ref{eq:proposed_constrained_flow_problem}). Results are shown for all combinations of network architecture (MLP network vs.\ Fourier features) with entropy viscosity formulations (linear $\nu^l_a$  vs.\ adaptive vanishing $\nu^a_a$)}
    \label{fig:lid_cavity_re2500_comp_v_prof}
\end{figure}

Figure~\ref{fig:lid_cavity_re2500_comp_v_prof} presents the mean and standard deviation (Std. band) of the predicted vertical velocity profiles, $v(x, 0.5)$, obtained using different formulations and four combinations of design choices.
In Figure~\ref{fig:lid_cavity_re2500_comp_v_prof}(a), which corresponds to the baseline formulation of the objective function, none of the combinations achieves fully accurate predictions. Among them, the Fourier-feature network with $\nu_a^a$, adaptive vanishing viscosity, yields results closest to the reference data. In contrast, the Fourier-feature network with linear entropy viscosity $\nu_a^l$, produces inconsistent mean trends and exhibits a wider deviation band.

\begin{figure}[!htb]
    \centering
    \includegraphics[width=0.9\linewidth]{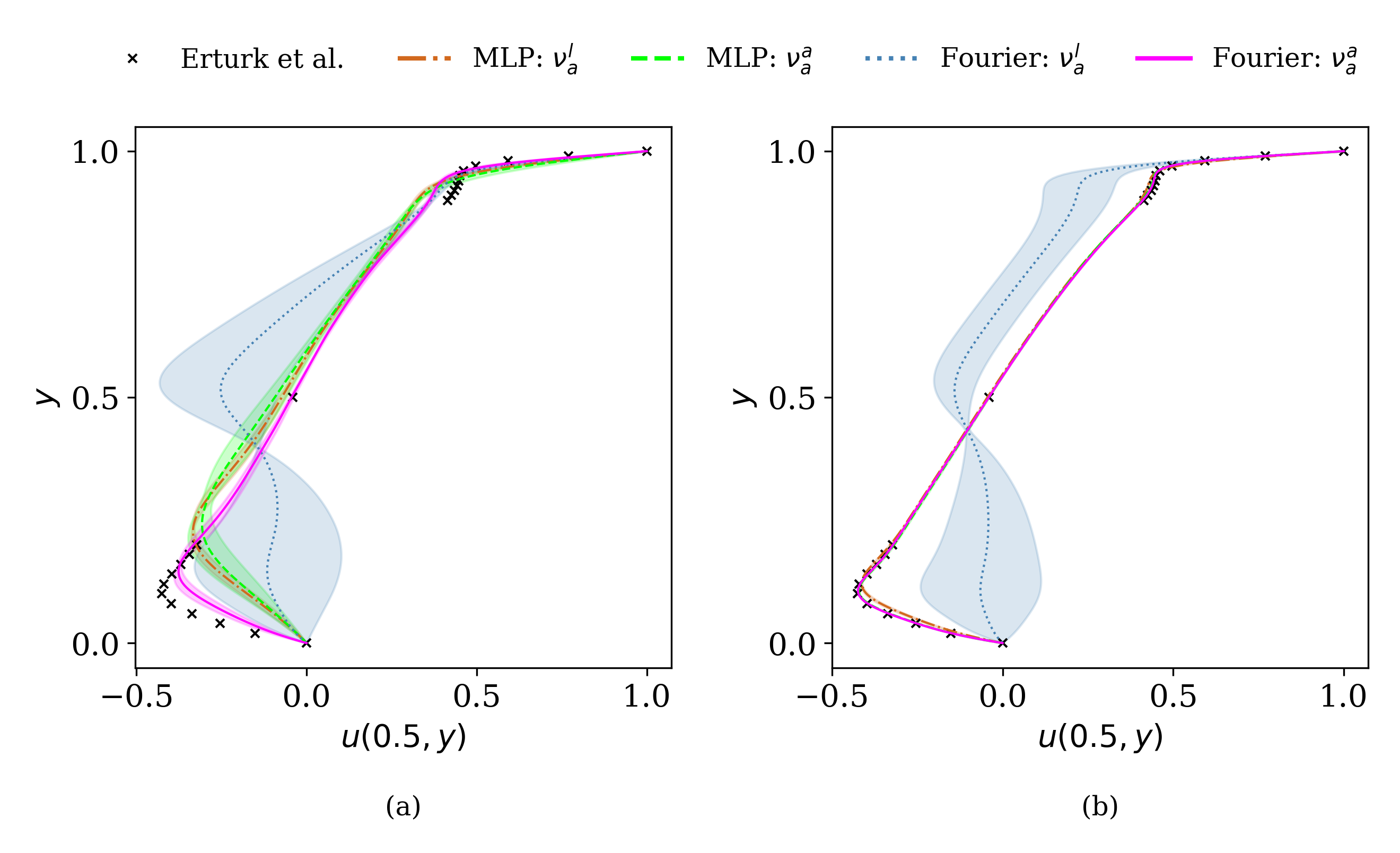}
    \caption{Predicted $u$-velocity along $x = 0.5$ for all configurations. See Fig. \ref{fig:lid_cavity_re2500_comp_v_prof}, for legend.}
    \label{fig:lid_cavity_re2500_comp_u_prof}
\end{figure}

When employing the proposed pressure-Poisson-based formulation, Figure~\ref{fig:lid_cavity_re2500_comp_v_prof}(b) shows notably improved prediction accuracy across configurations, except when Fourier-feature network is combined with the linear non-vanishing viscosity $\nu^l_a$. The Fourier-feature network with adaptive vanishing viscosity $\nu^a_a$ achieves the most accurate results, while the MLP network with $\nu^a_a$ also performs well, though a slight mismatch with the reference data persists.
The u-velocity profiles, $u(0.5, y)$ in Figure~\ref{fig:lid_cavity_re2500_comp_u_prof} exhibit the same trend.

\begin{figure}[!htb]
    \centering
    \includegraphics[width=0.9\linewidth]{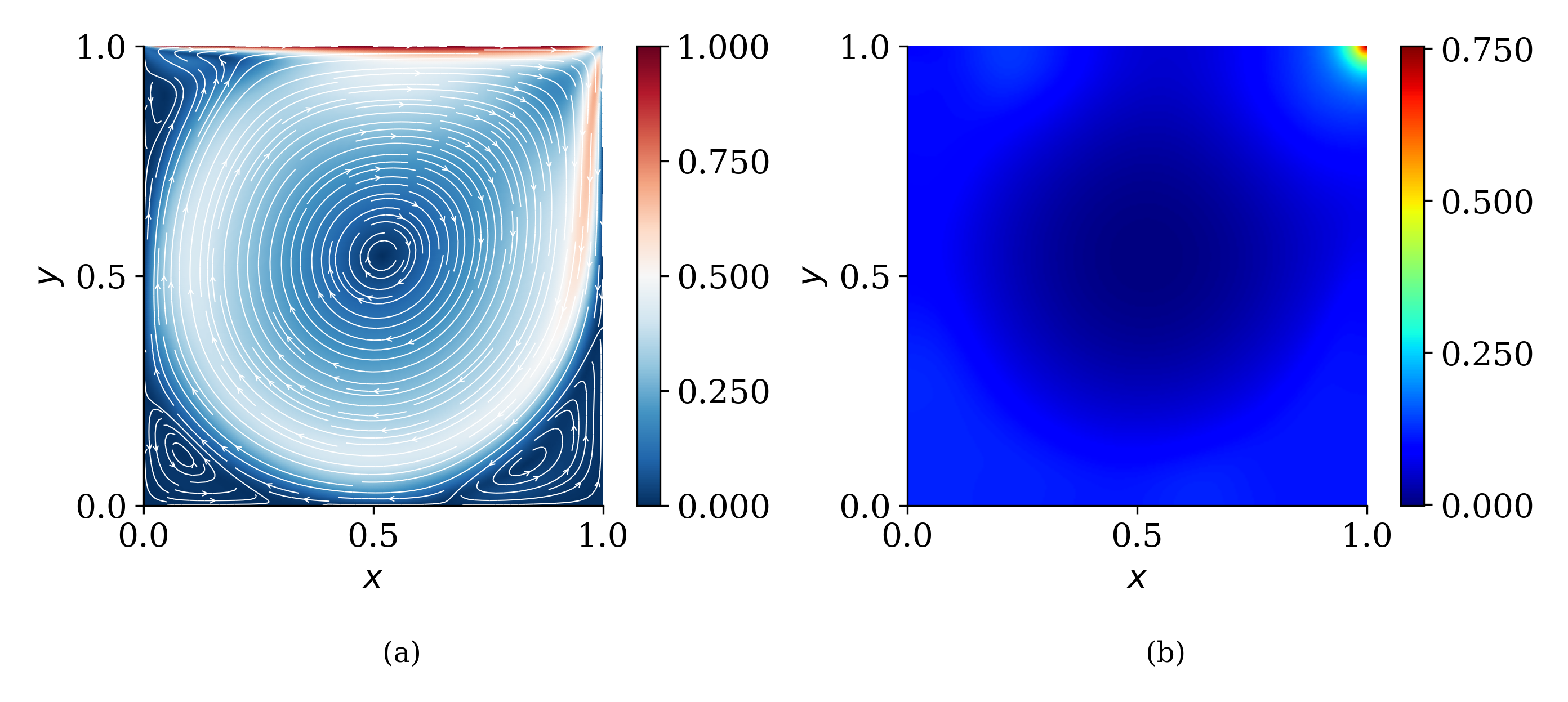}
    \caption{Lid-driven cavity flow at $Re = 2500$: (a) predicted velocity field with streamlines for the pressure-Poisson-based formulation with the Fourier-feature network and adaptive vanishing viscosity $\nu_a^a$, (b) the corresponding pressure field.}
    \label{fig:lid_cavity_re2500_contour}
\end{figure}

Figure~\ref{fig:lid_cavity_re2500_contour} shows the predicted velocity magnitude with streamlines and pressure distribution for the pressure-Poisson-based formulation with the Fourier-feature network and $\nu_a^a$.
In Fig.~\ref{fig:lid_cavity_re2500_contour}(a), the implementation successfully captures all four vortical structures with closed streamlines, particularly the small fourth vortex near the top-left corner.

\begin{figure}[!htb]
    \centering
    \includegraphics[width=1.0\linewidth]{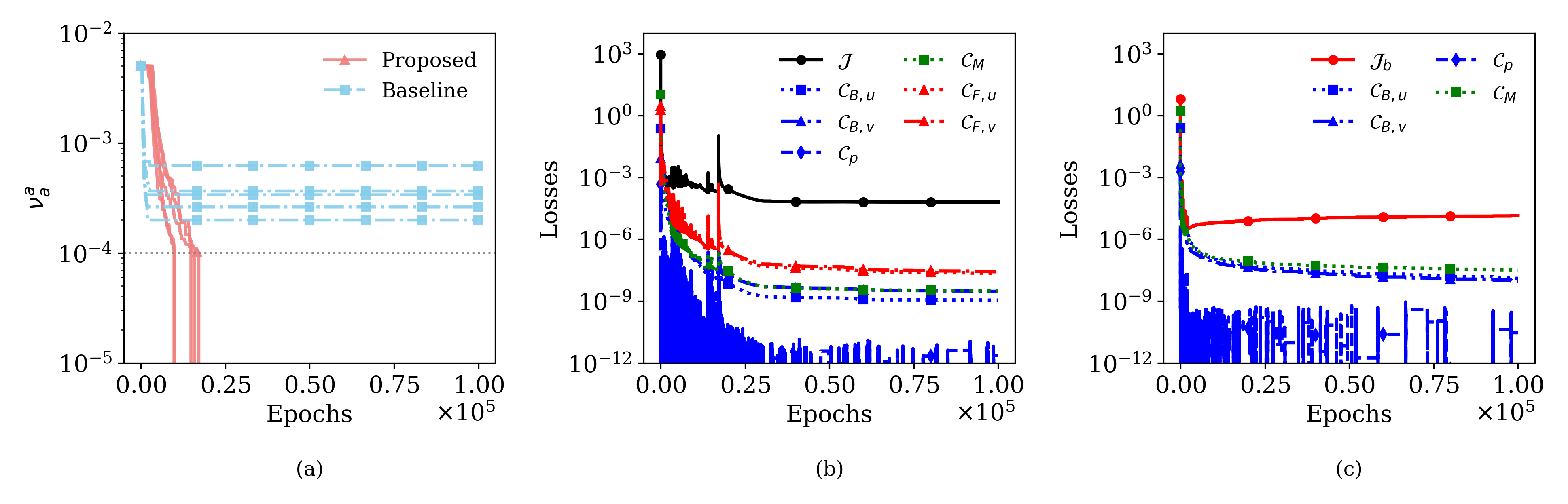}
    \caption{Lid-driven cavity flow at $Re = 2500$: comparison of the training behaviors between the proposed pressure-Poisson-based and the baseline objective formulations using the Fourier-feature network with adaptive vanishing viscosity $\nu_a^a$. (a) Evolution of $\nu_a^a$ over epochs for all trials, (b) evolution of the objective and constraint losses from one trial of the proposed formulation, and (c) evolution of the loss terms in the baseline formulation.}
    \label{fig:lid_cavity_re2500_training_evol}
\end{figure}

To analyze the training behavior of different formulations, we first focus on the configuration, Fourier-feature networks with adaptive vanishing viscosity $\nu_a^a$, that yields the closest agreement with the reference data even in the baseline cases, as shown in Fig.~\ref{fig:lid_cavity_re2500_comp_v_prof}(a) and Fig.~\ref{fig:lid_cavity_re2500_comp_u_prof}(a). Using this configuration, we compare the evolution of $\nu_a^a$ and the corresponding loss terms between the two formulations in Figure~\ref{fig:lid_cavity_re2500_training_evol}.

In Fig.~\ref{fig:lid_cavity_re2500_training_evol}(a), both formulations start from the same initial value of $\nu_a^{a(0)}=\tfrac{1}{200}$. In the five baseline trials, $\nu_a^a$ drops sharply during the initial stage but soon plateaus above the deactivation threshold of $10^{-4}$. In contrast, the proposed pressure-Poisson-based formulation maintains the initial value for several epochs before exhibiting a gradual decay, reaching the threshold before $2.5 \times 10^4$ epochs.
Figs.~\ref{fig:lid_cavity_re2500_training_evol}(b–c) show the corresponding objective and constraint losses from one trial of each formulation. The proposed pressure-Poisson-based formulation converges at approximately $4 \times 10^4$ epochs, with the boundary and divergence-free constraints stabilizing around $10^{-9}$, indicating near-exact satisfaction. The momentum losses converge to approximately $10^{-7}$, while the pressure-Poisson objective stabilizes near $10^{-4}$.
In contrast, the baseline objective converges more rapidly but stabilizes at a higher objective value (sum of the momentum losses) exceeding $10^{-6}$. Subsequent optimization fails to reduce this further, leaving both the maximum entropy constant and $\nu_a^a$ essentially unchanged.

\begin{figure}[!htb]
    \centering
    \includegraphics[width=1.0\linewidth]{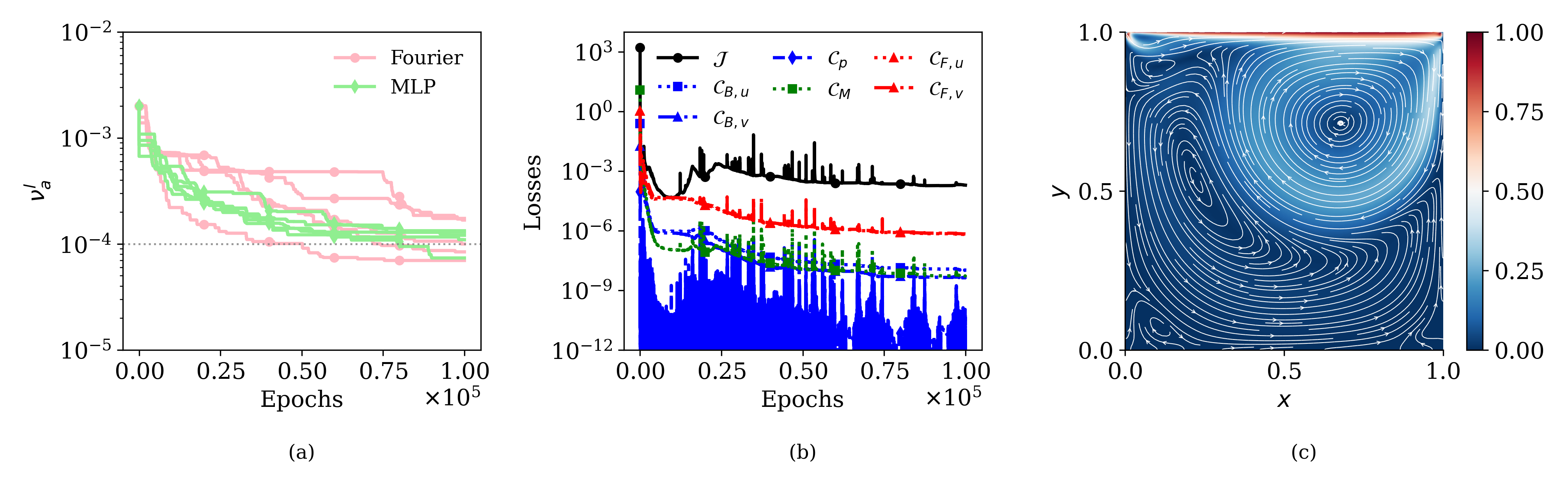}
    \caption{Lid-driven cavity flow at $Re = 2500$: (a) comparison of linear entropy viscosity $\nu_a^l$ evolution between the Fourier-feature and MLP networks under the pressure-Poisson-based formulation, (b) evolution of the objective and constraint losses and (c) contour visualization of the unexpected circulation pattern with streamlines from a Fourier-feature network trial. Color represents velocity magnitude.}
    \label{fig:lid_cavity_re2500_training_evol_fourier_linear}
\end{figure}

Another notable configuration is the Fourier-feature network equipped with the linear entropy viscosity $\nu_{a}^{l}$, which exhibits inconsistent performance across both formulations, as shown in Figs.~\ref{fig:lid_cavity_re2500_comp_v_prof} and~\ref{fig:lid_cavity_re2500_comp_u_prof}. 
To investigate this degraded behavior, Fig.~\ref{fig:lid_cavity_re2500_training_evol_fourier_linear} shows the evolution of $\nu_{a}^{l}$ under the pressure-Poisson-based formulation, together with the loss history and the predicted velocity field from a representative Fourier-feature network trial.

In Fig.~\ref{fig:lid_cavity_re2500_training_evol_fourier_linear}(a), the linear updates of $\nu_a^l$ decay much more slowly for both architectures, and few trials reach the deactivation threshold defined in the adaptive scheme, unlike those in Fig.~\ref{fig:lid_cavity_re2500_training_evol}(a). Furthermore, the Fourier-feature network exhibits pronounced inconsistency in the rate of decay compared to the MLP network.
The linear entropy viscosity does not vanish at convergence, which explains the slight mismatch in the velocity profiles obtained by the MLP network in Figs.~\ref{fig:lid_cavity_re2500_comp_v_prof}(b) and~\ref{fig:lid_cavity_re2500_comp_u_prof}(b), — the residual artificial diffusion due to non-vanishing linear entropy viscosity continues to degrade prediction accuracy even after training. However, this mechanism alone cannot account for the degraded performance observed in the Fourier-feature network.

Fig.~\ref{fig:lid_cavity_re2500_training_evol_fourier_linear}(b) shows highly fluctuating loss evolution over epochs from one of trials with Fourier-feature networks with linear entropy viscosity. Each spike in the loss curves corresponds to a decrease in $\nu_a^l$. The loss values at convergence are also higher than those observed with adaptive vanishing viscosity in Fig.~\ref{fig:lid_cavity_re2500_training_evol}(b), particularly for the momentum constraints, which stabilize around $10^{-6}$.

The corresponding predicted velocity field in Fig.~\ref{fig:lid_cavity_re2500_training_evol_fourier_linear}(c) reveals an unexpected circulation pattern in major disagreement with reference benchmark \citep{erturk_numerical_2005}. Velocity attains values that are very small across most of the domain except for a localized high-velocity region near the top-right corner. A similar solution with an unexpected circulation pattern in the cavity was also reported in \citet{SHUKLA2024117290} for Kolmogorov–Arnold networks (KAN) using a point-wise entropy viscosity.

\begin{figure}[!htb]
\centering
    \subfloat[]{\includegraphics[width=1.\textwidth]{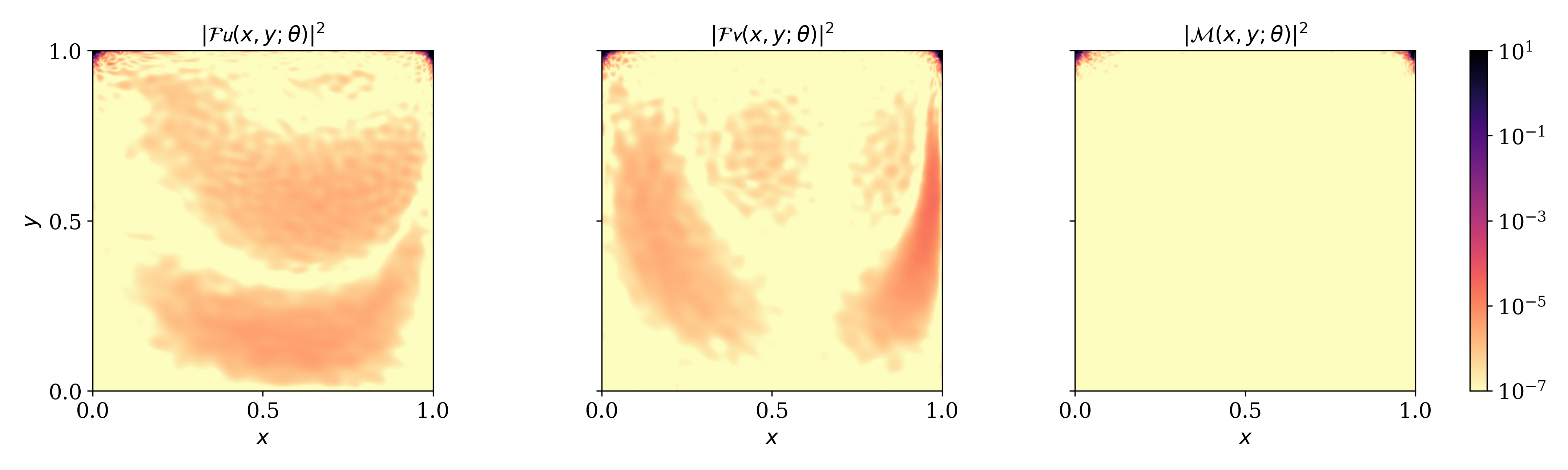}}
    \\
    \subfloat[]{\includegraphics[width=1.\textwidth]{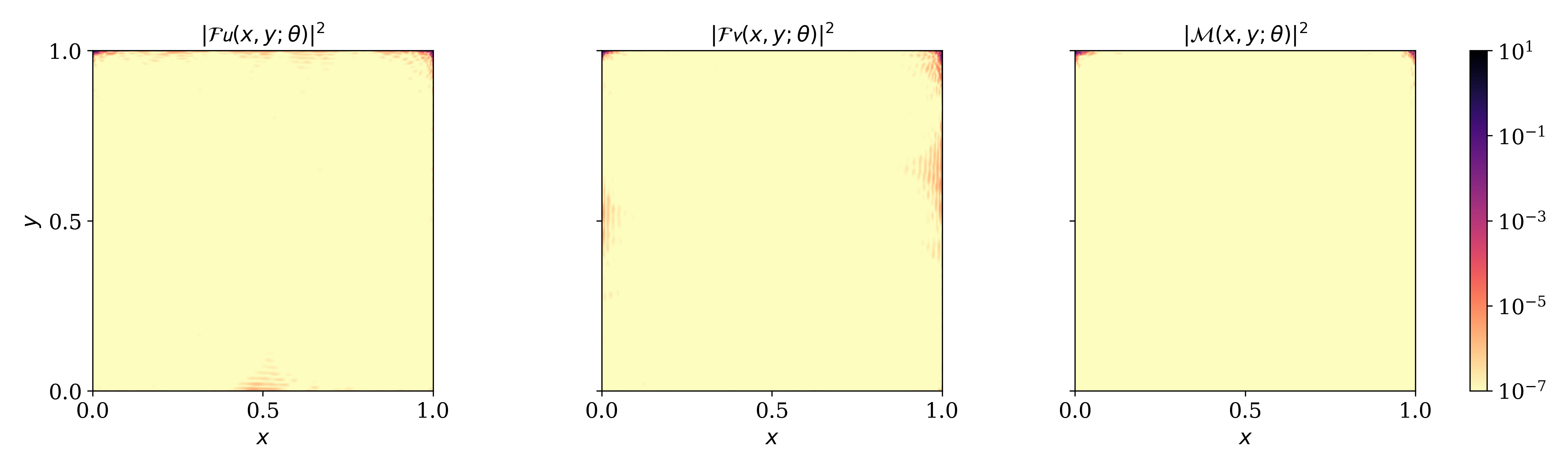}}
    \caption{Lid-driven cavity flow at $Re = 2500$: distribution of squared residuals of the momentum and mass conservation equations, evaluated on an unseen mesh for (a) the linear entropy viscosity $\nu_a^{l}$ and (b) the adaptive vanishing entropy viscosity $\nu_a^{a}$. Both cases use the proposed pressure-Poisson-based formulation with Fourier-feature networks, and residuals are computed from the modified governing equations that include the entropy viscosity.}
    \label{fig:lid_cavity_re2500_unseen_res_dis}
\end{figure}

To investigate the formation of this unexpected circulation pattern in the cavity, a $256 \times 256$ test mesh is used to examine the distribution of squared residuals of the momentum equations on unseen points in Fig.~\ref{fig:lid_cavity_re2500_unseen_res_dis}. Panel (a) shows the residuals from the adoption of linear entropy viscosity, whereas panel (b) shows the residuals due to adaptive vanishing viscosity. Residuals are evaluated from the momentum equations with the entropy viscosity terms included, At the final epoch, the linear strategy retains $\nu_a^{l} = 8.39 \times 10^{-5}$ (roughly one-fifth of the physical viscosity), which should have negligible impact on the global prediction, whereas $\nu_a^{a}$ vanishes. Despite smoothing the velocity discontinuities at the lid corners, Fig.~\ref{fig:lid_cavity_re2500_unseen_res_dis}(a) still exhibits residuals of extremely high magnitude near these singular points, and elevated momentum residuals persist throughout the interior.
The phenomena may be related to the low initial value of $\nu_a^l$ which scales as $\beta/Re$, and diminishes with increasing Reynolds number, potentially interacting unfavorably with the increased representational capacity of Fourier-feature networks. While this hypothesis warrants further investigation, it is beyond the scope of the present study.

Based on this ablation study --- with additional cases at $Re=5,000$ and $Re=7,500$ presented in Appendices A and B, respectively --- we adopt the pressure-Poisson-based formulation for all subsequent cases. 

The adaptive vanishing entropy viscosity $\nu^a_a$ is recommended only for high-$Re$ flows and is not employed in the lower-Reynolds-number examples that follow.

\subsection{Two-dimensional steady flow over a cylinder at $Re=40$}
\label{sec:steady_flow_cylinder}
We now demonstrate that the proposed pressure-Poisson-based formulation extends naturally to complex geometries with inflow and outflow boundary conditions. As a canonical example, we consider two-dimensional laminar flow over a cylinder at $Re=40$. This steady-state regime features two symmetric recirculation vortices behind the cylinder; the drag coefficient, separation angle, and wake length serve as validation metrics \citep{gautier2013_re40_cylinder}.

The rectangular computational domain contains a cylinder of unit diameter $D$ centered at the origin, with each boundary located $12D$ away from the cylinder surface. A total of 30,000 residual points are randomly sampled inside the domain, excluding those within the cylinder, along with 512 points on the cylinder surface and 1,024 points on each boundary. 

A constant-velocity Dirichlet boundary condition is prescribed at the inlet, where the components ($u_{in}$ and $v_{in}$) are determined from potential flow theory based on a mean unit inflow velocity. Symmetry (free-slip) conditions are applied on the lateral boundaries parallel to the inflow direction, enforcing a zero normal velocity ($v=0$) and a zero normal derivative of the tangential velocity ($u_y=0$). The pressure $p(x_{\min} + 1, 0) = 0$ is anchored at a single reference point located one diameter away from the inlet to prevent its unbounded growth.

Outflow boundary conditions for incompressible external flows have been extensively studied in conventional CFD \citep{gresho_1987_outflow, HASAN_2005_outflow, DONG_2014_outflow}, but these formulations do not transfer directly to the present optimization-based framework. In practice, we place the outlet sufficiently far from the region of interest to reduce sensitivity to the outflow treatment. To ensure global mass conservation, we borrow a technique from \citet{Blosch1993_outflow} that enforces equality between the incoming and outgoing mass flow rates. This is incorporated as an additional equality constraint alongside the outflow boundary conditions on primitive variables that will be introduced later.
\begin{equation}
\mathcal{C}_{\dot{m}}
= \left\|
\sum_{i=1}^{N_{\text{out}}} u_{out}(x_{\max}, y_i) \delta y_i
- \sum_{j=1}^{N_{\text{in}}} u_{in}(x_{\min}, y_j) \delta y_j
\right\|_2^2 = 0,
\end{equation}
where $N_{\text{out}}$ and $N_{\text{in}}$ denote the number of uniformly spaced sampling points on the outlet and inlet boundaries, respectively. $\delta y$ represents the effective discrete area for each collocation point on the boundary such that mass flow rate at the inlet matches the mass flow rate at the outlet. Note that density is constant due to the incompressible flow assumption.

To formulate these cases consistently, the same types of physical quantities are grouped into corresponding constraint terms. Accordingly, the resulting optimization problem takes the form:
\begin{equation}
    \begin{aligned}
        \min_{\theta} \mathcal{J}(\theta; \bm{x}), & \quad \text{subject to} \\
        \bm{\mathcal{C}}(\theta; \bm{x}) & = \big[\, \mathcal{C}_{F,u},\, \mathcal{C}_{F,v},\, \mathcal{C}_{M};\,
        \mathcal{C}_{B,u},\, \mathcal{C}_{B,v},\, \mathcal{C}_{p},\, \mathcal{C}_{B,u_y},
        \mathcal{C}_{\text{out}} \,\big]^{\!\top}
      = \bm{0},
    \end{aligned}
    \label{eq:steady_flow_cylinder_formulation}
\end{equation}
where $\mathcal{C}_{\text{out}} \subseteq \{\mathcal{C}_{\dot{m}},\, \mathcal{C}_{B,p_n},\, \mathcal{C}_{B,u_n},\, \mathcal{C}_{B,v_n}\}$ collects the outlet-treatment terms activated for a given ablation case, namely the mass flow rate balance and the Neumann conditions on pressure and velocity components, with $n$ denoting the direction normal to the outlet.

\begin{table}[t]
\centering
\footnotesize
\caption{Flow over a cylinder at $Re = 40$: ablation study on outflow boundary condition formulations for primitive variables at the outlet. Options include Neumann condition on pressure ($\partial p/\partial n = 0$), a global mass flow rate conservation ($\dot{m}$), and Neumann conditions on velocity components ($\partial u/\partial n = 0$, $\partial v/\partial n = 0$). \textit{Physical} denotes configurations whose solutions consistently reproduce the benchmark flow across repeated trials; \textit{Spurious} denotes configurations that converge consistently but to a non-physical solution.
%, with representative failure modes discussed in Section~\ref{sec:...}. 
The reported drag coefficient $C_D$ is given as mean $\pm$ standard deviation over five trials.}
\label{tab:ablation_outlet_cylinder}
\begin{tabular}{ccccc|lc}
\toprule
$\partial p/\partial n$ & $\dot{m}$\ & $\partial u/\partial n$ & $\partial v/\partial n$ & & Outcome & $C_D$ \\
\midrule
%\rowcolor{yellow!25}
\checkmark & \checkmark & --         & --         & & Physical  & $1.534 \pm 0.011$ \\
\checkmark & --         & \checkmark & \checkmark & & Spurious  & $0.272 \pm 0.094$ \\
\checkmark & \checkmark & --         & \checkmark & & Physical  & $1.544 \pm 0.007$ \\
\checkmark & \checkmark & \checkmark & --         & & Physical  & $1.533 \pm 0.002$ \\
\checkmark & \checkmark & \checkmark & \checkmark & & Physical  & $1.537 \pm 0.011$ \\
\midrule
--         & \checkmark & --         & --         & & Spurious & $0.235 \pm 0.046$ \\
--         & \checkmark & --         & \checkmark & & Spurious & $0.240 \pm 0.041$ \\

--         & \checkmark & \checkmark & --         & & Physical & $1.549 \pm 0.007$ \\

--         & \checkmark & \checkmark & \checkmark & & Physical & $1.544 \pm 0.004$\\
\bottomrule
\end{tabular}
\end{table}

Here, a simple MLP network with three hidden layers of 60 neurons each is trained for 20{,}000 epochs. Table~\ref{tab:ablation_outlet_cylinder} reports the ablation study on outflow boundary condition formulations for primitive variables at the outlet, with the drag coefficient $C_D$ given as the mean $\pm$ standard deviation across five independent trials, evaluated from 8192 equally spaced points on the cylinder. Among the configurations that include the pressure Neumann condition, only the one omitting the mass flow rate constraint consistently produces a spurious solution, yielding $\overline{C_D} < 0.2$, well below the benchmark value in Table~\ref{tab:flow_cylinder_re40_stat_comp}. All remaining configurations recover $\overline{C_D}$ around $1.5$, in close agreement with the reference value of $1.534 \pm 0.037$ reported as an average of several investigations \citep{gautier2013cylinder}.

Notably, a physical solution can be recovered without any pressure boundary condition, provided global mass flow rate conservation is enforced at the outlet as a constraint. Comparing Neumann conditions of both $u-$ and $v-$ components of the velocity, $\partial u / \partial n = 0$ proves essential for convergence, whereas $\partial v / \partial n = 0$ plays a markedly weaker role.

\begin{figure}[!htb]
\centering
    \subfloat[]{\includegraphics[width=1.\textwidth]{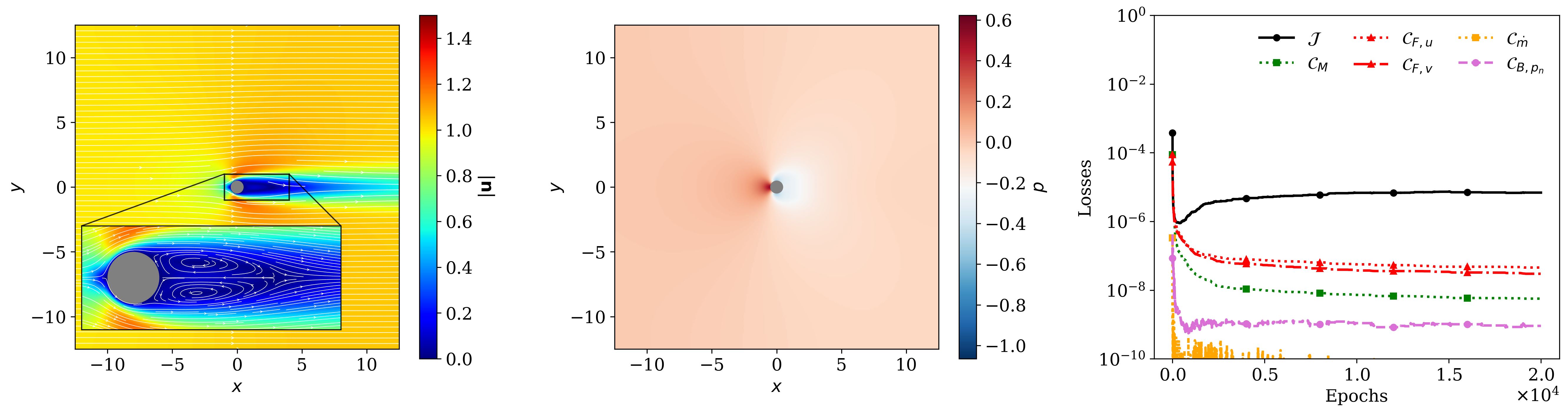}} 
    \\
    \subfloat[]{\includegraphics[width=1.\textwidth]{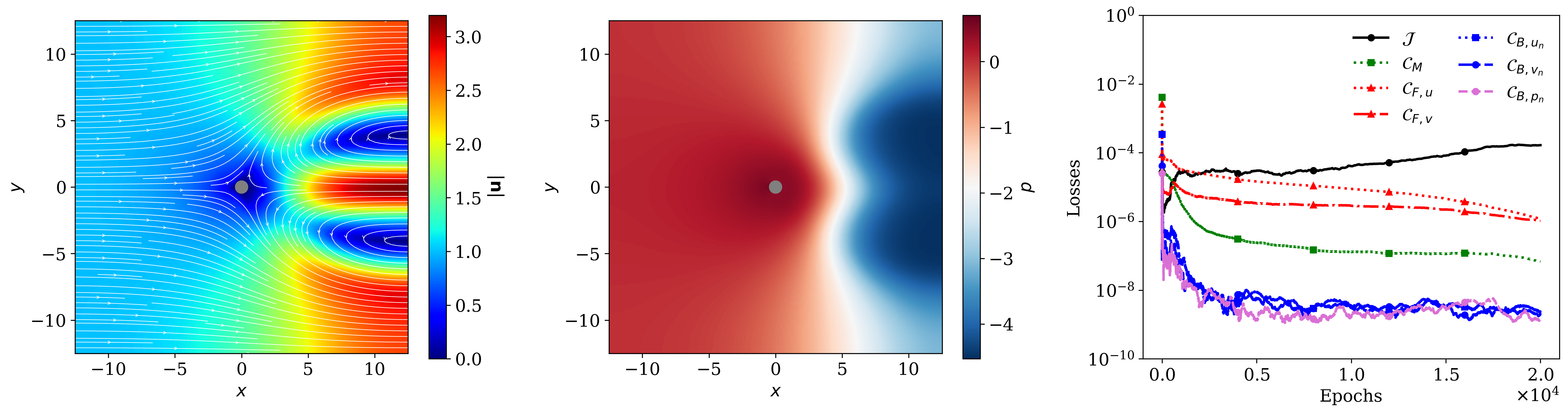}}
    \caption{Flow over a cylinder at $Re = 40$: predictions and loss histories for two outlet configurations sharing $\partial p/\partial n = 0$, supplemented by (a) the global mass flow rate conservation constraint, yielding a physical solution, and (b) $\partial u/\partial n = \partial v/\partial n = 0$, yielding a spurious solution. Columns (left to right): velocity magnitude, pressure, loss history.}
    \label{fig:flow_cylinder_re40_success_failure_comparison}
\end{figure}

\begin{figure}[!htb]
\centering
    \subfloat[]{\includegraphics[width=1.\textwidth]{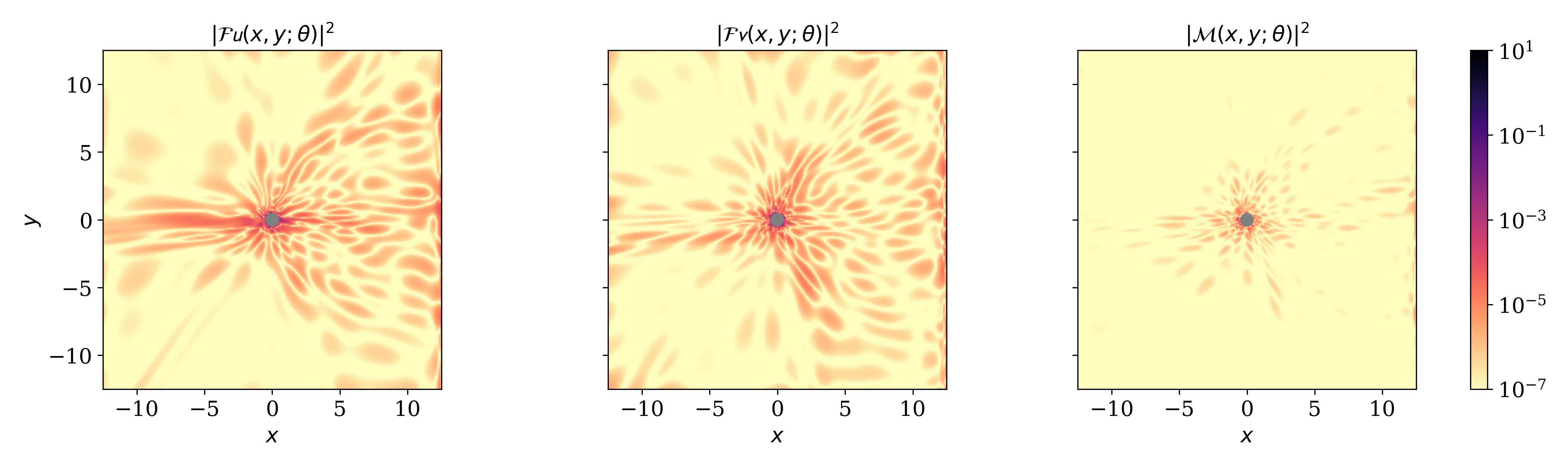}} 
    \\
    \subfloat[]{\includegraphics[width=1.\textwidth]{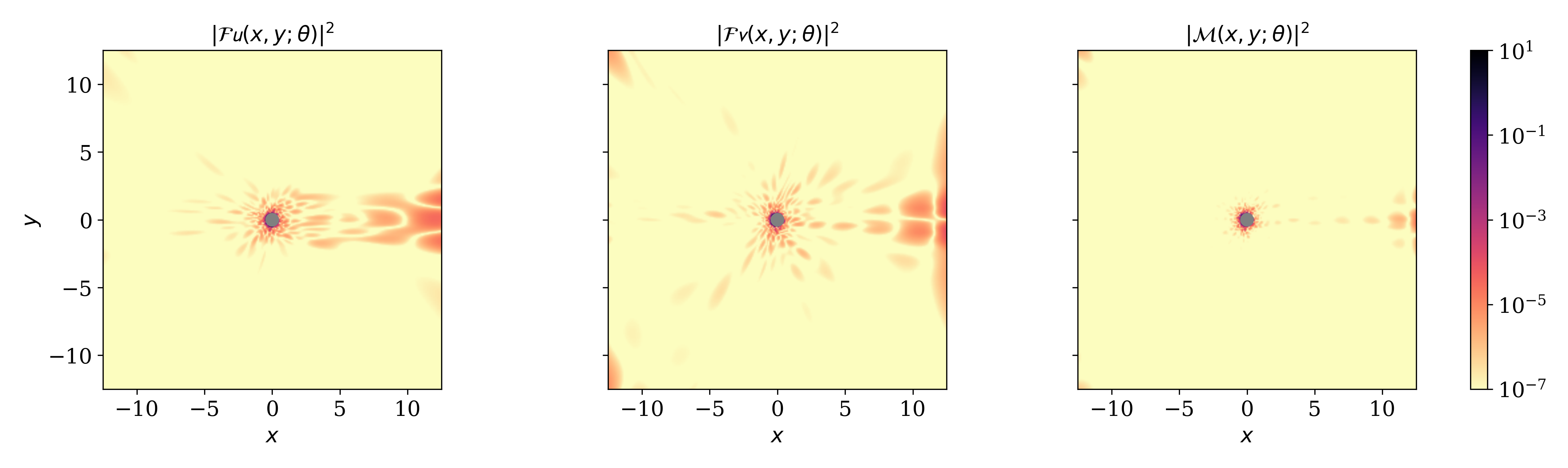}}
    \\
    \subfloat[]{\includegraphics[width=1.\textwidth]{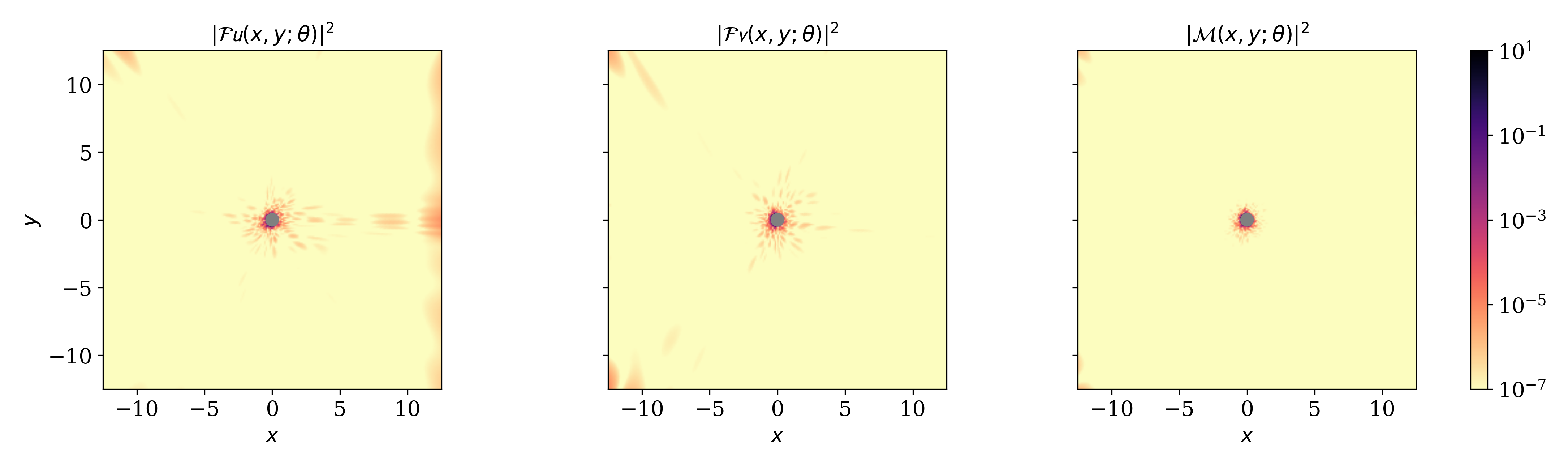}} 
    \caption{Steady flow over a cylinder at $Re = 40$: distribution of squared residuals of the momentum and continuity equations, evaluated on unseen points during training, for three outlet configurations: (a) $\partial p/\partial n = \partial u/\partial n = \partial v/\partial n = 0$. Solution is spurious, (b) the mass flow rate conservation constraint with $\partial u/\partial n = \partial v/\partial n = 0$. Solution is physical, and (c) the mass flow rate conservation constraint with $\partial p/\partial n = 0$. Solution is physical.}
    \label{fig:flow_cylinder_re40_unseen_sqrd_res_comp}
\end{figure}

Figure~\ref{fig:flow_cylinder_re40_success_failure_comparison} compares the predicted velocity and pressure fields and the corresponding loss histories for two outlet configurations producing physical and spurious outcomes. In Fig.~\ref{fig:flow_cylinder_re40_success_failure_comparison}(a), corresponding to the combination of the mass flow rate constraint $\mathcal{C}_{\dot{m}}$ and $\partial p/\partial n = 0$, the predicted fields follow the expected physical behavior, with clear flow obstruction by the cylinder; the zoomed-in view further reveals two symmetric recirculation vortices in the near wake. The loss evolution shows convergence of the objective, residual of the $u-$ and $v-$ momentum equations, and outlet constraints, apart from a brief rise in the objective during early training. Notably, the outlet constraints converge first, followed by the divergence-free constraint to near $10^{-8}$ after roughly 5{,}000 epochs, while the momentum constraints settle around $10^{-7}$.

Fig.~\ref{fig:flow_cylinder_re40_success_failure_comparison}(b) illustrates a representative spurious prediction, obtained by implementing pure Neumann outlet conditions on both velocity components and pressure. A reverse flow with abnormally high velocity magnitude appears near the outlet, accompanied by a disturbed pressure field extending into the far field. This behavior suggests a violation of global mass conservation, even though the differential divergence-free condition is enforced inside the domain. The loss evolution corroborates the failure: the objective continues to grow gradually beyond $10^{-4}$ throughout training, while the momentum constraints decrease only slowly and remain above $10^{-6}$, well above the levels achieved in panel~(a).

To further investigate the performance difference between physical and spurious solutions, we present the distribution of squared residuals of the momentum and continuity equations in Figure~\ref{fig:flow_cylinder_re40_unseen_sqrd_res_comp}, evaluated on $256\times256$ test points unseen during network training. \textcolor{black}{Strictly speaking, solution accuracy cannot be judged from residual levels alone due to the nonlinearity of the Navier-Stokes equations. Therefore, the present visual comparison of squared residuals provides a qualitative assessment of different outlet conditions.} 

Three outlet configurations are assessed in Fig.~\ref{fig:flow_cylinder_re40_unseen_sqrd_res_comp} : pure Neumann conditions for both velocity components and pressure, which produced a spurious solution, and two configurations employing the global mass flow rate constraint $\mathcal{C}_{\dot{m}}$ combined with either a Neumann condition on the velocity components or on pressure, both of which produced a physical solution. As seen in Fig.~\ref{fig:flow_cylinder_re40_unseen_sqrd_res_comp}(a), the spurious solution exhibits elevated momentum residuals in both the upstream and downstream regions, with partial violations of the divergence-free condition. The two physical solutions incorporating $\mathcal{C}_{\dot{m}}$ also exhibit notably different behaviors: the $\partial u/\partial n = \partial v/\partial n = 0$ case in Fig.~\ref{fig:flow_cylinder_re40_unseen_sqrd_res_comp}(b) displays larger momentum residuals in the wake, while the $\partial p/\partial n = 0$ case in Fig.~\ref{fig:flow_cylinder_re40_unseen_sqrd_res_comp}(c) maintains comparatively smaller residuals there, despite some visible $u-$momentum residuals near the outlet. This behavior is physically consistent: the wake disturbance induced by the cylinder causes the flow at the portion of the outlet directly downstream of the cylinder to deviate from the zero-gradient condition imposed on the velocity components, thereby degrading solution quality in that region.

\begin{table}[t]
\centering
\footnotesize
\caption{Flow over a cylinder at $Re = 40$: mean and standard deviation of drag coefficient $C_D$, separation angle $\theta_s$, and normalized wake length $L_w/D$, obtained from three outlet configurations with $\mathcal{C}_{\dot{m}}$, compared against conventional numerical results.}
\label{tab:flow_cylinder_re40_stat_comp}
\begin{tabular}{l|ccc}
\toprule
 & $C_D$ & $\theta_s$ & $L_w/D$ \\
\midrule
Dennis and Chang \citep{dennis1970_cylinder}  & $1.52$ & $126.2\degree$ & $2.35$\\
Fornberg \citep{fornberg1980_cylinder}  & $1.50$ & $124.4\degree$ & $2.24$ \\
He and Doolen \citep{he1997_cylinder}  & $1.50$ & $127.2\degree$ & $2.25$ \\
Ye et al. \citep{ye1999_cylinder} & $1.52$ &  & $2.27$ \\
Calhoun \citep{calhoun2002cartesian} & $1.62$ & $125.8\degree$ & $2.18$ \\
Linnick and Fasel \citep{linnick2005high} & $1.54$ & $126.4\degree$ & $2.28$ \\
Ding et al. \citep{ding2007applications} & $1.58$ & $127.2\degree$ & $2.35$ \\
Taira and Colonius~\cite{taira2007immersed} & $1.54$ & $126.3\degree$ & $2.30$ \\
Patil and Lakshmisha~\cite{patil2009finite} & $1.56$ & $127.3\degree$ & $2.14$ \\
Bouchon et al.~\cite{bouchon2012second} & $1.50$ & $126.6\degree$ & $2.26$ \\
Gautier et al. \citep{gautier2013cylinder} & $1.49$ & $126.4\degree$ & $2.24$ \\
Mean of various works from \citep{gautier2013cylinder} & $1.534 \pm 0.037$ & $126.38\degree \pm
0.84\degree$ & $2.260 \pm  0.059$ \\
\midrule
$\boldsymbol{\mathcal{C}_{\text{out}} = [\mathcal{C}_{\dot{m}}, \mathcal{C}_{B,p_n}]}$ & $\boldsymbol{1.534 \pm 0.011}$ & $\boldsymbol{126.35\degree \pm 0.84\degree}$ & $\boldsymbol{2.222 \pm 0.036}$ 
\\
$\mathcal{C}_{\text{out}} = [\mathcal{C}_{\dot{m}}, \mathcal{C}_{B,u_n}, \mathcal{C}_{B,v_n}]$ & $1.544 \pm 0.004$ & $126.67\degree \pm 0.26\degree$ & $2.221 \pm 0.012$
\\
$\mathcal{C}_{\text{out}} = [\mathcal{C}_{\dot{m}}, \mathcal{C}_{B,p_n}, \mathcal{C}_{B,u_n}, \mathcal{C}_{B,v_n}]$ & $1.537 \pm 0.011$ & $126.71\degree \pm 1.25\degree$ & $2.228 \pm 0.018$
\\
\bottomrule
\end{tabular}
\end{table}

Table~\ref{tab:flow_cylinder_re40_stat_comp} compares the drag coefficient $C_D$, separation angle $\theta_s$, and normalized wake length $L_w/D$ against values reported in the literature. Results from the proposed pressure-Poisson-based formulation adopting three different outlet configurations, each employing the mass flow rate constraint $\mathcal{C}_{\dot{m}}$. All configurations yield close agreement with the mean values reported in the literature. Among them, a single outlet Neumann condition for pressure combined with $\mathcal{C}_{\dot{m}}$ achieves the best agreement with reference values of $C_D$ and $\theta_s$, with only the wake length showing a slightly larger deviation. Based on this finding and the residual analysis in Fig.~\ref{fig:flow_cylinder_re40_unseen_sqrd_res_comp}, a zero Neumann condition for pressure paired with global mass flow rate conservation $\mathcal{C}_{\dot{m}}$ is identified as the preferred outlet configuration and is adopted for the unsteady flow over a cylinder at $Re = 100$ in Section \ref{sec:unsteady_cylinder}.

\subsection{\textcolor{black}{Two-dimensional steady flow over a backward-facing step at $Re=800$}}
\textcolor{black}{To further investigate the outlet-boundary treatment, we consider the canonical flow over a backward-facing step (BFS) problem, which has been adopted as a benchmark problem in several studies to assess numerical treatment of outlet conditions \citep{Kim1985, gartling1990, Blosch1993_outflow, erturk2008}. 
The flow behind the step exhibits a complex structure: the shear layer detaches at the step corner and reattaches on the lower wall, while an adverse pressure gradient drives a second, longer recirculation zone on the opposite wall, and the extent of both regions varies strongly with Reynolds number and with the streamwise extent of the computational domain. 
In contrast to the cylinder wake, the fluid leaves the domain as a developing, strongly sheared channel flow that has not yet recovered its fully developed profile, so any spurious upstream influence of the outlet is registered directly in the predicted separation and reattachment locations. 
}

\textcolor{black}{
Following Gartling~\cite{gartling1990}, we consider a rectangular domain with an expansion ratio of two, $\Omega=\{(x,y)|x\in[0, L_{out}], y\in[-H/2, H/2]\}$, so that the downstream channel has height $H=1$m while the step height and the inlet channel height are both $H/2$. 
A fully developed plane Poiseuille profile $u(y) = 24\,y\,(H/2 - y)$, $v = 0$, is prescribed at the inlet plane, giving a mean inflow velocity $\bar{u} = 1.0$m/s, and no-slip conditions are imposed on all solid walls. 
The Reynolds number is defined as $Re = \bar{u} H / \nu = 800$, which permits a direct comparison of separation and reattachment lengths with published studies. We normalize the length and velocity with $H$ and $\bar{u}$, respectively.
}

\textcolor{black}{
We place our outlet boundary at $L_{out}=15$, which is half the channel length used in~\cite{gartling1990}. At $L_{out}=15$, the outlet lies within the region where the flow is still recovering toward its fully developed profile and the boundary treatment is therefore subject to a stringent test. 
We use the Fourier-feature network with 3 hidden layers and 60 neurons per layer.
The domain is sampled with 60{,}000 interior collocation points, together with 256 points on the inlet, 256 on the step face, 512 on the outlet, and 2048 on each of the upper and lower walls.
The pressure field is anchored at an upstream location $p(0.25, 0.25) = 0$.
The pressure field is anchored near the inlet, by imposing $p(0.5,0.25)=0$.
}

\begin{figure}[!htb]
\centering
    \includegraphics[width=1.\textwidth]{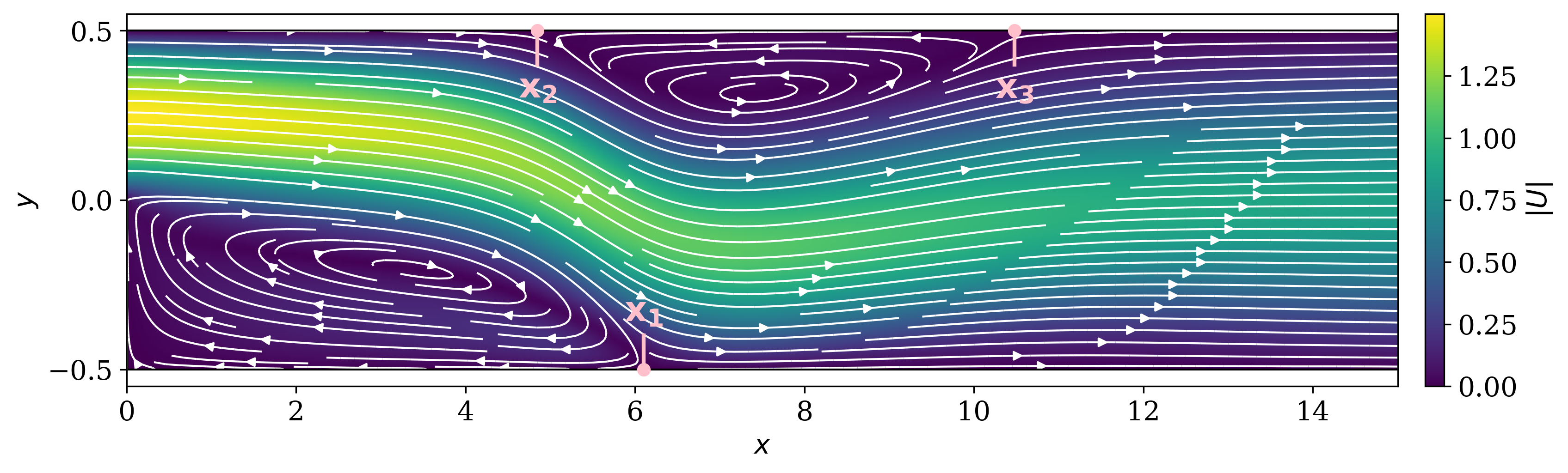} 
    \caption{\textcolor{black}{Flow over a backward facing step at $Re=800$: predicted velocity magnitude with streamlines, obtained with the zero Neumann conditions on the pressure coupled with a global mass flow rate conservation treatment ($\boldsymbol{\mathcal{C}_{\text{out}} = [\mathcal{C}_{\dot{m}}, \mathcal{C}_{B,p_n}]}$) at the outlet of $L_{out}=15$. The streamwise locations $x_1, x_2, x_3$ mark the detachment and reattachment points. Figure not drawn to scale.}}
    \label{fig:bfs_vel_contour}
\end{figure}

\textcolor{black}{
Figure~\ref{fig:bfs_vel_contour} shows the predicted distribution of velocity magnitude with the streamlines, using the outlet treatment of pressure Neumann and global mass flow rate conservation. 
The parabolic inlet profile separates at the step corner and forms a large recirculation eddy on the lower wall, whose reattachment point lies at $x_1 \approx6.1$, i.e.\ twelve step heights downstream of the step.
The resulting shear layer deflects toward the lower wall and creates a second recirculation zone on the upper wall, which separates at $x_2\approx4.9$ and reattaches at $x_3\approx10.5$.
}

\begin{figure}[!htb]
\centering
    \includegraphics[width=0.95\textwidth]{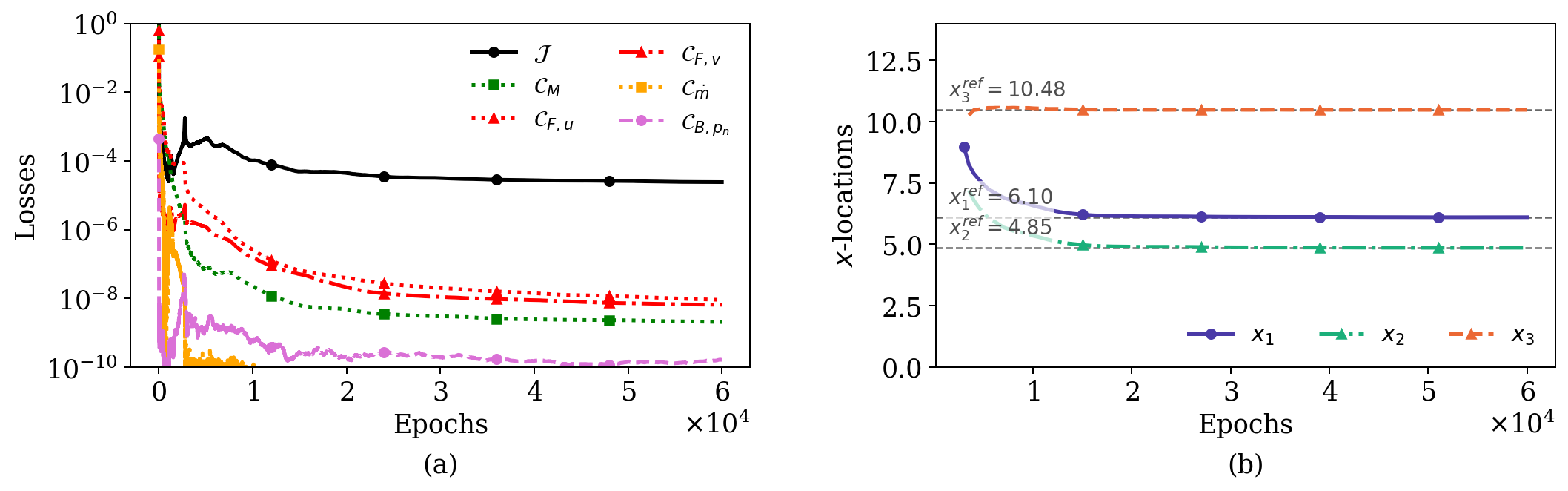} 
    \caption{\textcolor{black}{Flow over a backward-facing step at $Re=800$: (a) Convergence history of the objective $\mathcal{J}$, the mass and momentum constraints $\mathcal{C}_{M}$, $\mathcal{C}_{F,u}$, $\mathcal{C}_{F,v}$, and the two outlet constraints: the global mass flow rate residual $\mathcal{C}_{\dot{m}}$ and the pressure Neumann residual $\mathcal{C}_{B,p_n}$, with $L_{out}=15$. (b) Evolution of the reattachment point of the lower-wall eddy $x_1$ and the separation and reattachment points of the upper-wall eddy $x_2$, $x_3$; dashed horitonzal lines mark the reference values of Gartling~\cite{gartling1990}.}}
    \label{fig:bfs_loss_x_hist}
\end{figure}

\textcolor{black}{
The convergence history is summarized in Figure~\ref{fig:bfs_loss_x_hist}. Panel~(a) shows that all loss terms converges smoothly after an initial transient of roughly $3\times10^{3}$ epochs, during which the objective $\mathcal{J}$ passes through a local maximum.
The mass and momentum constraints settle below $10^{-8}$, and the two outlet constraints are satisfied to tighter tolerance. 
Panel~(b) tracks the three keypoints that characterize the separated flow. 
All three are identified within the first $10^{4}$ epochs and then converge close to the benchmark values of Gartling~\cite{gartling1990}, before $3\times10^4$ epochs. 
Convergence of keypoint locations indicates that the outlet boundary placed at $L_{out}=15$ does not adversely affect the upstream separation zones.
}

\textcolor{black}{
Table~\ref{tab:bfs_keypoints} compares the three keypoint locations with values reported in the literature as compiled by~\citet{erturk2008}. The values are rescaled to the present normalization. 
Our predicted locations $x_1=6.106 \pm 0.018$, $x_2=4.863 \pm 0.019$ and $x_3=10.486 \pm 0.009$ agree with the benchmark solution of~\citet{gartling1990} to within $0.10\%$, $0.27\%$ and $0.06\%$ respectively, and lie within one standard deviation of the mean across all nine studies.
The standard deviation of our predictions is smaller than the one of the published values by roughly an order of magnitude, indicating that the separation and reattachment locations are recovered repeatably rather than as the outcome of a single favorable run. 
These locations are known to be sensitive to the streamwise extent of the domain, to the outflow specification~\citep{gartling1990}, and even to the length of the inlet channel upstream of the step~\citep{barton1997,erturk2008};
Therefore, the accurate recovery of the separation and reattachment locations on a domain truncated to $L_{out}=15$ provides further evidence that the outlet treatment, combining a pressure Neumann condition with global mass-flow-rate conservation, does not adversely affect the upstream flow physics in any significant way.
}

\begin{table}[t]
\footnotesize
\centering
\caption{\textcolor{black}{Flow over a backward-facing step at $Re=800$: comparison of the reattachment point of the lower-wall recirculation $x_1$ and the separation and reattachment points of the upper-wall recirculation $x_2$ and $x_3$. The values are scaled by the height of the downstream channel $H=1$.}}
\label{tab:bfs_keypoints}
\begin{tabular}{l|ccc}
\toprule
 & $x_1$ & $x_2$ & $x_3$ \\
\midrule
Gartling~\citep{gartling1990}          & 6.100 & 4.850 & 10.480 \\
Barton~\citep{barton1997}              & 5.755 & 4.570 & 10.330 \\
Kim and Moin~\citep{Kim1985}           & 5.950 & --    & --     \\
Guj and Stella~\citep{guj1988}         & 6.025 & 4.850 & 10.150 \\
Gresho et al.~\citep{gresho1993}       & 6.100 & 4.860 & 10.490 \\
Keskar and Lyn~\citep{keskar1999}      & 6.095 & 4.855 & 10.480 \\
Grigoriev and Dargush~\citep{grigoriev1999} & 6.090 & 4.850 & 10.470 \\
Rogers and Kwak~\citep{rogers1991}     & 5.740 & --    & --     \\
Erturk~\citep{erturk2008}              & 5.917 & 4.738 & 10.277 \\
Mean of various works from~\citep{erturk2008} & $5.975 \pm 0.145$ & $4.796 \pm 0.109$ & $10.382 \pm 0.133$ \\
\midrule
$\boldsymbol{\mathcal{C}_{\text{out}} = [\mathcal{C}_{\dot{m}}, \mathcal{C}_{B,p_n}]}$ & $6.106 \pm 0.018$ & $4.863 \pm 0.019$ & $10.486 \pm 0.009$ \\
\bottomrule
\end{tabular}
\end{table}

\begin{figure}[!htb]
\centering
    \includegraphics[width=1.\textwidth]{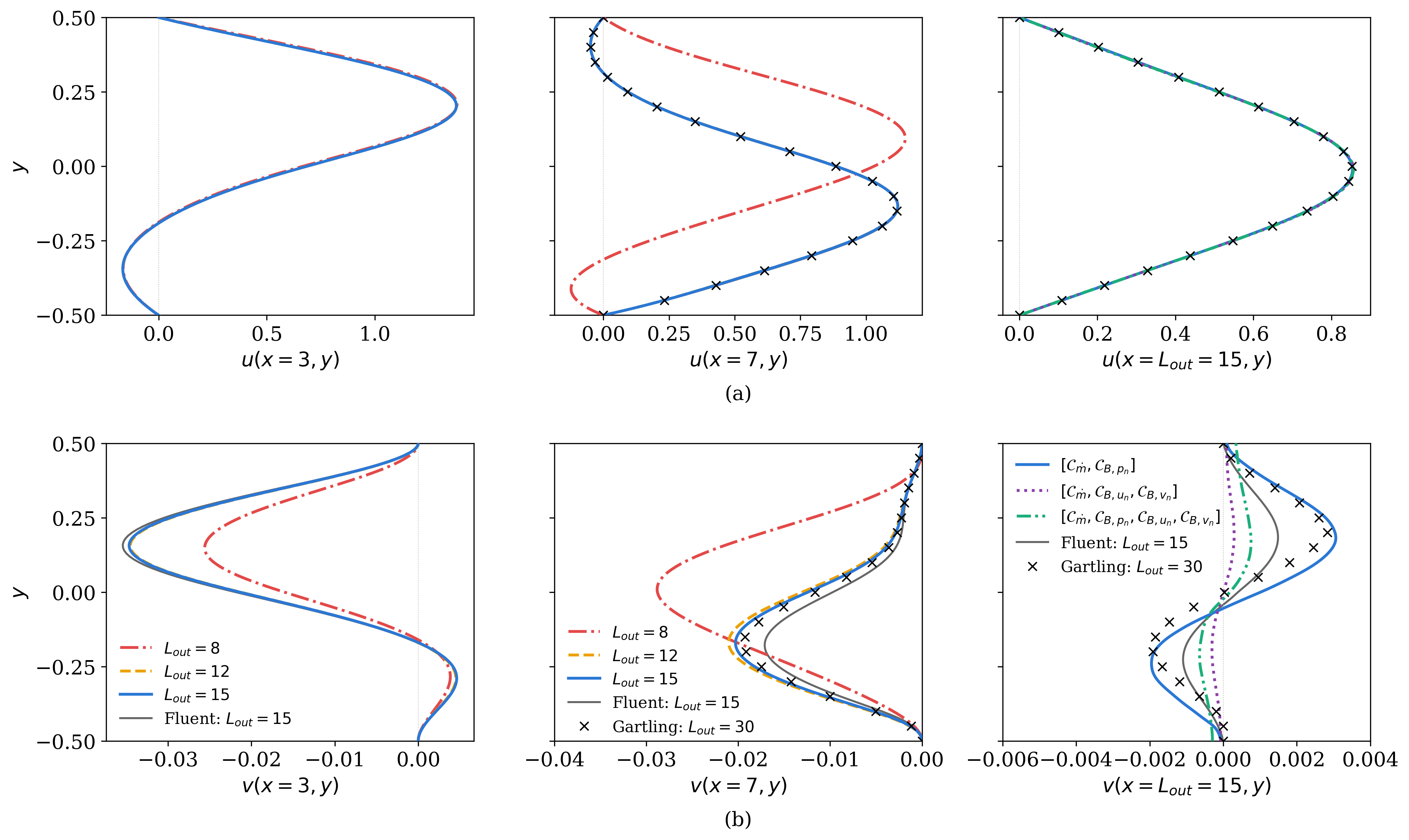}
    \caption{\textcolor{black}{Flow over a backward-facing step at $Re=800$: cross-channel profiles of (a) $u$ and (b) $v$ velocity at $x=3$, $x=7$ and the outlet plane $x=L_{out}=15$, compared with an Ansys Fluent solution using $L_{out}=15$ and with the benchmark data of~\citet{gartling1990} using $L_{out}=30$. The first two columns compare outlets placed at $L_{out} \in \{8, 12, 15\}$ under the pressure Neumann and global mass flow rate conservation; the third column compares three combinations of outlet constraints at the outlet plane itself: mass flow rate with pressure Neumann $[\mathcal{C}_{\dot m}, \mathcal{C}_{B,p_n}]$, mass flow rate with velocity Neumann $[\mathcal{C}_{\dot m}, \mathcal{C}_{B,u_n}, \mathcal{C}_{B,v_n}]$, and all three combined.}}
    \label{fig:bfs_prof_comp}
\end{figure}

\textcolor{black}{
Figure~\ref{fig:bfs_prof_comp} assesses the influence of outlet boundary treatment on the velocity profiles at three locations along the channel, including the outlet plane itself.
The $u$-velocity profiles in Fig.~\ref{fig:bfs_prof_comp}(a) are mutually consistent at every station, with the sole exception of $x=7$ obtained with a shorter channel $L_{out}=8$, and the same discrepancy is evident in the corresponding $v$-velocity profiles in Fig.~\ref{fig:bfs_prof_comp}(b).
Note that with $L_{out}=8$, the outlet plane cuts through the recirculation zone on the upper wall resulting in the displacement of the high-speed zone toward the inlet.}
\textcolor{black}{
As no benchmark profile is available at $x=3$, we performed an additional simulation using Ansys Fluent software on the same computational domain ($L_{out}=15$), with quadrilateral elements of size $0.01$. We imposed the ``outflow'' boundary condition available in Ansys Fluent, which also imposes a zero Neumann condition on pressure coupled with a global mass flow rate conservation.
The comparison shows that the present predictions approach the reference $v$-profile of~\citet{gartling1990} progressively as the outlet is moved downstream, and that the $L_{out}=15$ result agrees with it more closely than the Fluent solution obtained on the same domain.
Therefore, the interior stations are genuinely sensitive to the outlet location, and the global mass conservation and pressure Neumann treatment recovers them accurately as long as the outlet lies far from the final reattachment point.
}

\textcolor{black}{
The profiles at $x=L_{out}=15$ in the last column of Fig.~\ref{fig:bfs_prof_comp} compare three combinations of Neumann conditions imposed together with the global mass flow rate conservation at the outlet plane.
Since the reference solution of \citet{gartling1990} at this station was computed on a domain with $L_{out}=30$, the comparison provides a direct assessment of how well each outlet treatment reproduces the developing flow on a truncated domain that half the length of the domain used in the reference solution.
All three combinations, along with the Ansys Fluent solution, recover the same $u$-profile, in excellent agreement with the reference and close to the fully developed parabolic shape; it is the $v$-component that discriminates between them.
The pressure Neumann condition reproduces the extrema (about $+3\times10^{-3}$ and $-2\times10^{-3}$) of the reference $v$-profile, with a slight overprediction of the positive peak.
Imposing the Neumann condition on the velocity components instead drives $v$ towards zero, while combining all three constraints retains the correct sign structure but suppresses the amplitude by roughly a factor of four; the Fluent solution recovers approximately half the reference amplitude.
Although $v$ is smaller than $u$ by around two orders of magnitude at this station, it does show that the pressure Neumann condition admits the residual transverse motion associated with the developing flow, whereas a velocity Neumann condition enforces a parallel state that the flow has not yet attained.}

\textcolor{black}{
Based on the results of the present study and the flow-over-cylinder case at $Re=40$, we recommend the use of a zero-Neumann pressure boundary condition combined with a global mass-conservation constraint, whereby the total mass flow rate entering the domain is required to match the total mass flow rate leaving it. The effectiveness of this approach is consistent with common CFD practice, although a notable advantage is that no additional boundary conditions need to be prescribed for the velocity components.
}

\subsection{Three-dimensional unsteady Beltrami flow}
\label{sec:beltrami}
In this section, we consider a spatio-temporal problem with analytical solutions: the three-dimensional Beltrami flow \citep{ethier1994exact}. This problem has been used as a benchmark in NSFnets \citep{jin_nsfnets_2021}, where the authors show that the vorticity-velocity formulation outperforms the primitive-variable formulation when prescribed weights are used to balance the boundary and initial condition terms in the composite loss. Their results also benefit from network depth: with 50 neurons each, 7 hidden layers perform slightly better than 4. Here, we adopt the same $4\times50$ MLP network, two-stage optimization strategy (Adam for 10{,}000 epochs followed by L-BFGS for 30,000 epochs), and number of residual points, 10{,}000, distributed in the spatio-temporal domain $\mathcal{U}=\{(x,y,z,t)\,|\,-1\leq x,y,z\leq 1,\ 0\leq t\leq 1\}$. Random sampling is applied entirely throughout, with 1{,}000 points per boundary face and 2{,}000 points for the initial condition.

Similar to NSFnets, we obtain the initial conditions and Dirichlet boundary conditions on the velocity and pressure fields from the analytical solution~\citep{ethier1994exact}, 
\begin{equation}
\begin{aligned}
u(x,y,z,t) &= -a\left[e^{ax}\sin(ay+dz) + e^{az}\cos(ax+dy)\right]e^{-d^{2}t}, \\
v(x,y,z,t) &= -a\left[e^{ay}\sin(az+dx) + e^{ax}\cos(ay+dz)\right]e^{-d^{2}t}, \\
w(x,y,z,t) &= -a\left[e^{az}\sin(ax+dy) + e^{ay}\cos(az+dx)\right]e^{-d^{2}t}, \\
p(x,y,z,t) &= -\tfrac{1}{2}a^{2}\Big[e^{2ax} + e^{2ay} + e^{2az} \\
&\qquad + 2\sin(ax+dy)\cos(az+dx),e^{a(y+z)} \\
&\qquad + 2\sin(ay+dz)\cos(ax+dy),e^{a(z+x)} \\
&\qquad + 2\sin(az+dx)\cos(ay+dz),e^{a(x+y)}\Big]e^{-2d^{2}t},
\end{aligned}
\label{eq:beltrami}
\end{equation}
with $a=d=1$. The Reynolds number is set to 1 to match the governing equations 
in~\citet{ethier1994exact}. Thus, the corresponding pressure-Poisson-based formulation is
\begin{equation}
    \begin{aligned}
        \min_{\theta} \mathcal{J}(\theta; \bm{x}, t), & \quad \text{subject to} \\
        \bm{\mathcal{C}}(\theta; \bm{x}, t) & = \big[\, \mathcal{C}_{F,u},\, \mathcal{C}_{F,v},\, \mathcal{C}_{F,w},\,\mathcal{C}_{M},\,
        \mathcal{C}_{B,u},\, \mathcal{C}_{B,v},\, \mathcal{C}_{B,w},\, \mathcal{C}_{B,p}, 
        \mathcal{C}_{I,u},\, \mathcal{C}_{I,v},\, \mathcal{C}_{I,w},\, \mathcal{C}_{I,p} \big]^{\!\top}
      = \bm{0}.
    \end{aligned}
    \label{eq:beltrami_flow_formulation}
\end{equation}
For evaluation, collocation points are obtained from a uniform spatial grid $101\times101\times101$ for nine time instants from $t=0$ to $t=1$ with a time step size of $\Delta t = 0.125$.

\begin{table}[t]
\centering
\footnotesize
\caption{Unsteady 3D Beltrami flow: statistical performance in terms of temporal mean relative $\ell^2$ errors. The NSFnet results are the mean of five independent trials as reported in~\citep{jin_nsfnets_2021}, using a $7\times50$ network with the vorticity-velocity (VV) and velocity-pressure (VP) formulations. Our proposed pressure-Poisson-based  formulation~\eqref{eq:beltrami_flow_formulation} uses a smaller $4\times50$ network.}
\label{tab:beltrami_l2_comparison}
\begin{tabular}{lrrr}
\toprule
 & NSFnet (VV) & NSFnet (VP) & \textbf{current work} (Eq.~\ref{eq:beltrami_flow_formulation}) \\
\midrule
$\overline{\mathcal{E}_r}(\hat{u},u)$ & $(2.38 \pm 0.33) \times 10^{-4}$ & $(7.31 \pm 1.49) \times 10^{-4}$ & $\bm{(5.37 \pm 0.48) \times 10^{-5}}$ \\
$\overline{\mathcal{E}_r}(\hat{v},v)$ & $(2.38 \pm 0.20) \times 10^{-4}$ & $(1.016 \pm 0.191) \times 10^{-3}$ & $\bm{(5.23 \pm 0.47) \times 10^{-5}}$ \\
$\overline{\mathcal{E}_r}(\hat{w},w)$ & $(2.41 \pm 0.16) \times 10^{-4}$ & $(9.68 \pm 2.26) \times 10^{-4}$ & $\bm{(5.07 \pm 0.45) \times 10^{-5}}$ \\
$\overline{\mathcal{E}_r}(\hat{p},p)$ & n/a & $(8.91 \pm 1.74) \times 10^{-2}$ & $\bm{(1.06 \pm 0.01) \times 10^{-4}}$ \\
\bottomrule
\end{tabular}
\end{table}

Table~\ref{tab:beltrami_l2_comparison} compares the average over five trials and the corresponding standard deviation of the temporal mean relative $\ell^2$ errors between the NSFnets predictions~\citep{jin_nsfnets_2021} and our pressure-Poisson-based  formulation~\eqref{eq:beltrami_flow_formulation}. The NSFnets study adopts both the velocity-pressure (VP) and velocity-vorticity (VV) formulations of the Navier-Stokes equations. The VV formulation by design does not produce the pressure field, whereas the VP formulation does. Here, we present the best results of NSFnets obtained by the deeper $7\times50$ network, while the shallower $4\times50$ network, which we adopt, was reported to underperform~\citep{jin_nsfnets_2021}. Despite this less expressive architecture, our proposed method achieves roughly an order-of-magnitude error reduction relative to the NSFnets VV formulation. Since both our method and NSFnets VP formulation use velocity-pressure variables, the latter is the more relevant comparison. Owing to the pressure-Poisson-based formulation, our pressure predictions are two orders of magnitude more accurate, with velocity components showing one to two orders of magnitude improvement. Overall, our predictions surpass both NSFnets formulations across all reported quantities.

\begin{figure}[!htb]
\centering
    \includegraphics[width=0.7\textwidth]{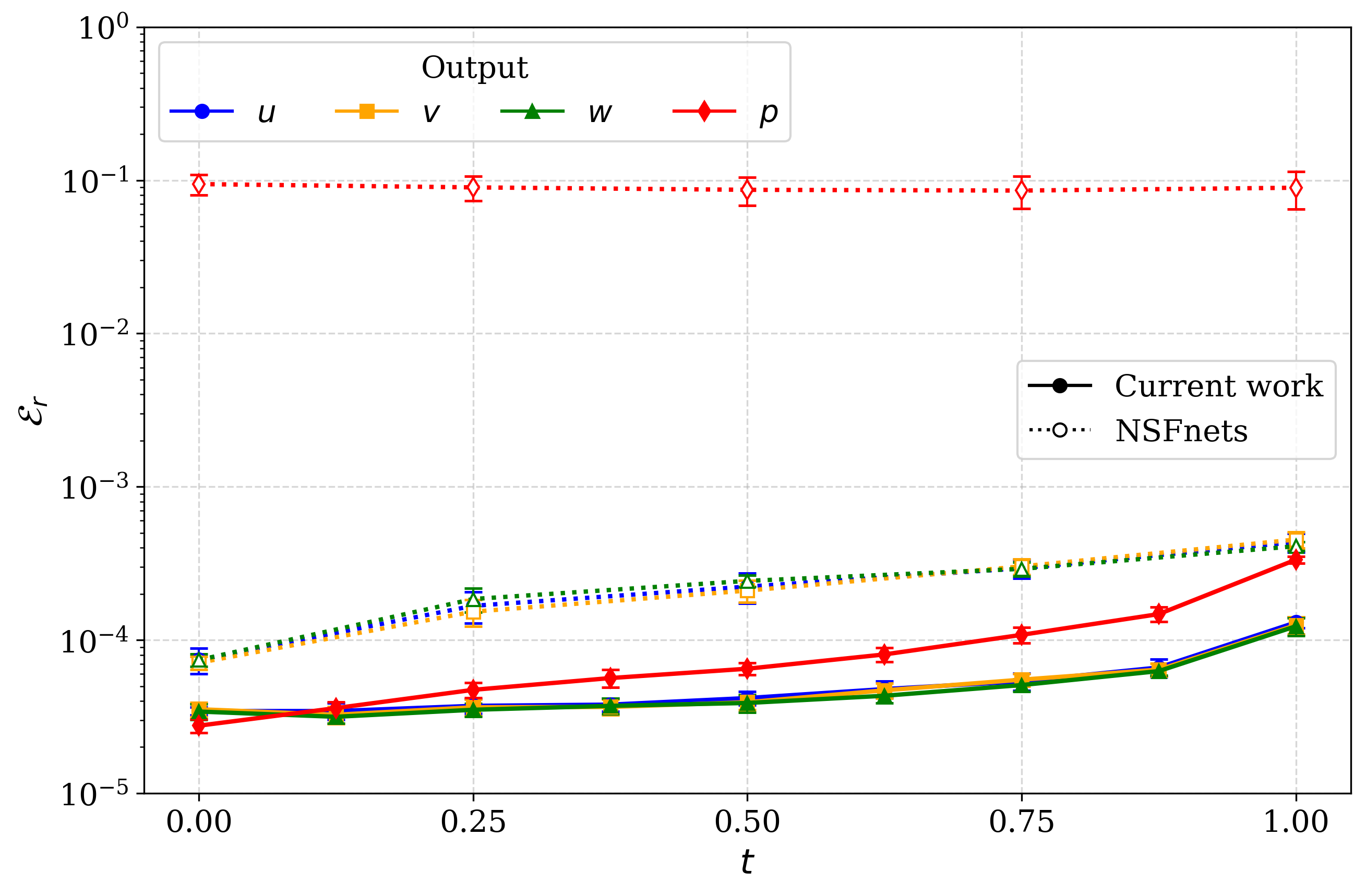} 
    \caption{Unsteady 3D Beltrami flow: temporal evolution of the mean and standard deviation of the relative $\ell^2$ errors for velocity components and pressure, compared against the NSFnets, where error values for velocity are from the VV formulation and pressure from the VP formulations of NSFnets ~\citep{jin_nsfnets_2021} (dotted lines with open markers).}
    \label{fig:beltrami_l2_time_comp}
\end{figure}

\begin{figure}[!htb]
\centering
    \includegraphics[width=1.\textwidth]{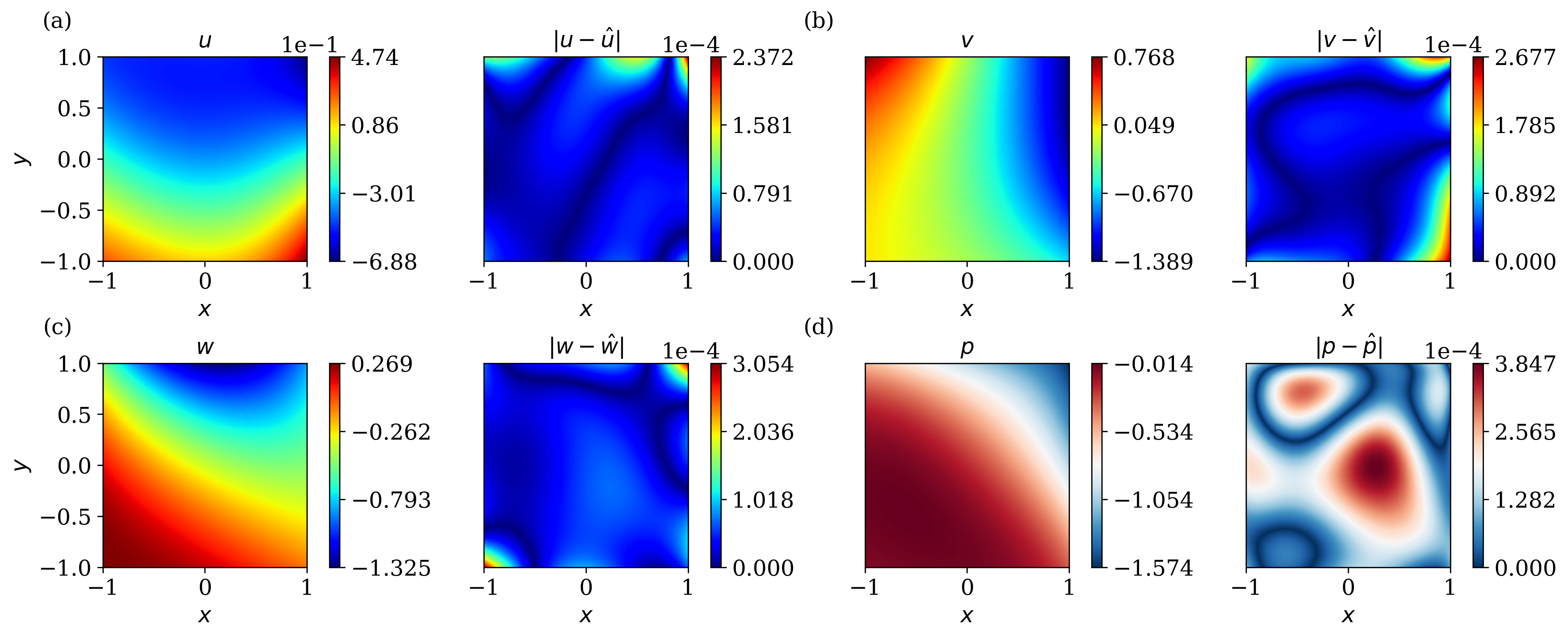} 
    \caption{Unsteady 3D Beltrami flow solution (exact) and absolute error distribution at $t=1.00$ on the plane $z=0$. (a) u-velocity, (b) v-velocity, (c) w-velocity, (d) pressure.}
    \label{fig:beltrami_pred_t1_z0}
\end{figure}

Figure~\ref{fig:beltrami_l2_time_comp} shows the temporal evolution of the relative $\ell^2$ errors for each velocity components and pressure. Overall our predictions exhibit much lower error levels than NSFnets. Both our predictions and NSFnets show an increasing trend of the error with time, pointing to the difficulty of long-time evolution with a spatio-temporal formulation. To address this issue, we adopt a discrete time formulation in the next unsteady flow example. 

A snapshot of the exact solutions at $t=1$ on the plane $z=0$ is shown in Figure~\ref{fig:beltrami_pred_t1_z0}, along with the absolute error distribution for the velocity components and pressure. While the maximum error values are comparable across the four panels, the pressure field in Fig.~\ref{fig:beltrami_pred_t1_z0}(d) shows a relatively larger error region in the interior of the spatial domain.

\subsection{Two-dimensional unsteady flow over a cylinder at $Re=100$}\label{sec:unsteady_cylinder}
The steady cylinder wake loses its stability at $Re \approx 47-49$ beyond which the attached vortices shed periodically, forming the well-known von K{\'a}rm{\'a}n street. The wake remains two-dimensional up to $Re\approx 140-194$ \citep{Williamson1996}. We consider impulsively started flow over cylinder at $Re=100$ (based on free-stream velocity and cylinder diameter).  This problem is particularly challenging for PINNs, as vortex shedding must arise spontaneously from a uniform inflow velocity and zero initial velocity in the domain without external perturbation.

As demonstrated for the 3D Beltrami flow, errors in spatio-temporal formulations accumulate with increasing time horizon. Since capturing periodic vortex shedding requires simulating over many shedding cycles, we instead adopt a discrete-time formulation and approximate the unsteady term in Eq.~\eqref{eq:uns_incompress_mom} using a second-order backward difference formula (BDF2) \citep{suli2003introduction}. The time derivative of the velocity field $\bm{u}$ in Eq.~\eqref{eq:uns_incompress_mom} is approximated at time level $t^{n+1}$ as:
\begin{equation}
    \left.\frac{\partial \bm{u}}{\partial t}\right|^{n+1} \approx \frac{3\bm{u}^{n+1} - 4\bm{u}^{n} + \bm{u}^{n-1}}{2\,\Delta t},
\end{equation}
where the time step size is set to $\Delta t=0.25$, and $\bm{u}^{n}$, $\bm{u}^{n-1}$ denote the values of $\bm{u}$ at the two previous time levels. Each time level is represented by a dedicated network, whose output is carried forward as the time integration advances. At time instant $t^{n+1}$, an individual network represents the velocity and pressure field
\begin{equation}
    [\hat{u}, \hat{v}, \hat{p}]^{\top}\!\big|_{t=t^{n+1}} = f_{\theta^{n+1}}(x, y),
\end{equation}
with the solution from previous time levels entering the training as known history terms.
In that regard, training a network with a discrete time advancement can be viewed as an implicit scheme.  Since BDF2 requires two prior solutions, the first time step (from $t^{0}$ to $t^{1}$) is performed using a first-order backward Euler scheme. Because prior states enter naturally through the BDF2 stencil as history terms, no initial condition constraint needs to be explicitly enforced, and the pressure-Poisson-based formulation~\eqref{eq:steady_flow_cylinder_formulation} from the steady case is retained unchanged.

Based on the outcome of our ablation study on outflow boundary conditions, we enforce a global mass flow rate conservation constraint and a zero Neumann condition on pressure on the outlet boundary, $\mathcal{C}_{\text{out}} = \{ \mathcal{C}_{\dot{m}}, \mathcal{C}_{B,p_n} \}$. We emphasize that a pressure boundary condition is enforced only at the outlet; all other boundaries carry velocity conditions.    

\begin{figure}[!htb]
\centering
    \includegraphics[width=1.\textwidth]{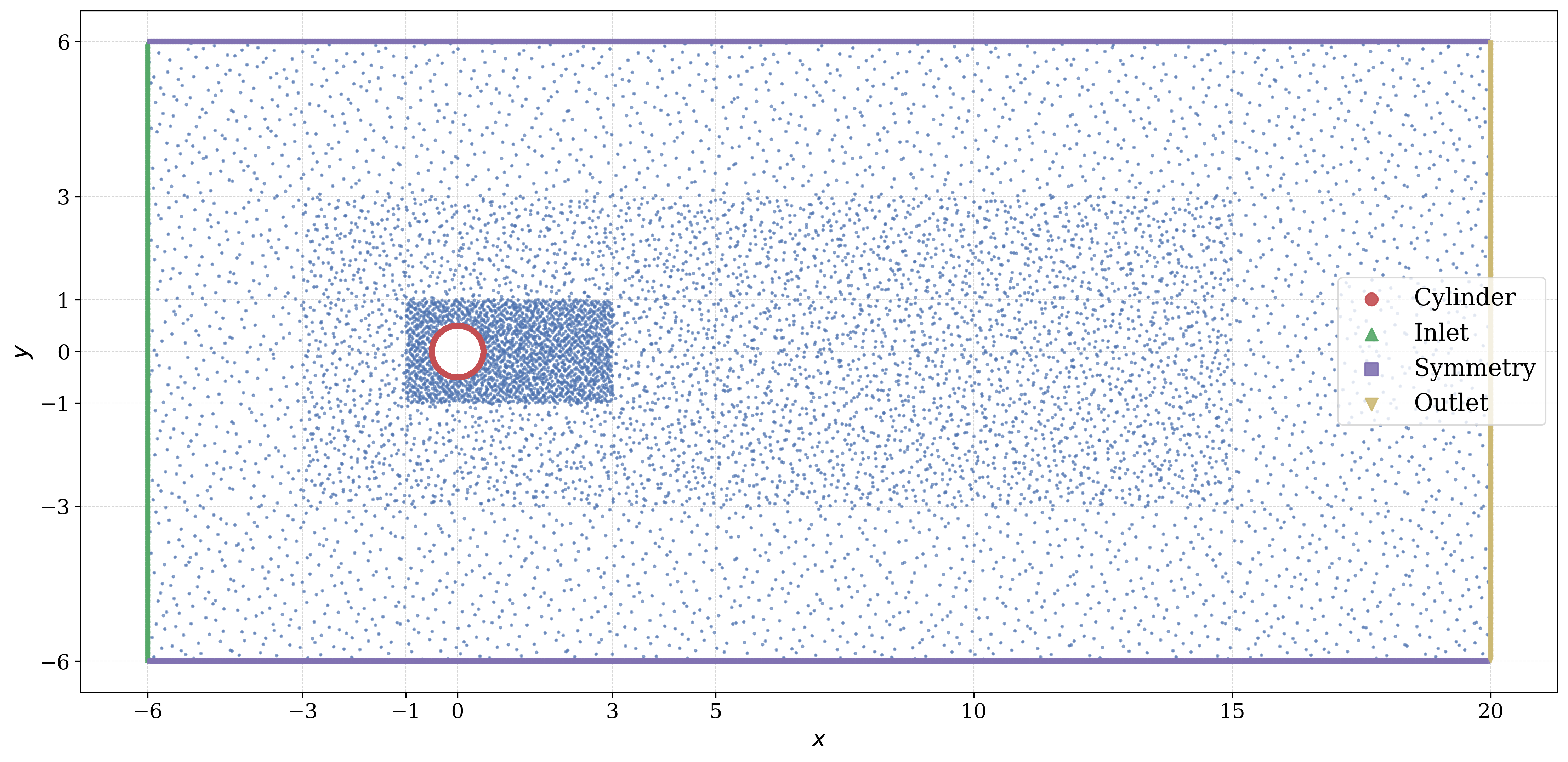} 
    \caption{Unsteady flow over a cylinder at $Re=100$: hierarchical three-level sampling of residual points, progressively refined from the global domain to the wake subdomain $\Omega_w$ and the near-cylinder subdomain $\Omega_c$, with boundary points shown on the cylinder surface, inlet, side walls, and outlet. Boundary conditions on domain boundaries are also shown.}
    \label{fig:unsteady_cylinder_point_distri}
\end{figure}

The spatial domain is defined as $\Omega = [-6, 20] \times [-6, 6]$, following \citep{HAUKE19981, Calderer2010}. As shown in Fig.~\ref{fig:unsteady_cylinder_point_distri}, we adopt a three-level hierarchical sampling strategy to resolve sharp velocity gradients near the cylinder and vortex shedding in the wake without inflating the total residual-point budget. The points are distributed progressively: 
$5000$ over the global domain $\Omega$, $3000$ over the intermediate wake 
subdomain $\Omega_w = [-3,15] \times [-3,3]$, and $2000$ over the 
near-cylinder subdomain $\Omega_c = [-1,3] \times [-1,1]$. On the boundaries, $1000$ points are placed along the cylinder surface, and $500$ points are placed on each of the far-field boundaries (inlet, outlet, symmetry boundaries on the top and bottom, see Fig.~\ref{fig:unsteady_cylinder_point_distri}).

To enhance the approximation capacity for the periodic shedding phenomenon, we consider a Fourier-feature network with 4 hidden layers and 60 neurons per layer. However, the implementation shows that this network is prone to overfitting during the early transient stage, when the initially zero velocity field has not yet been fully convected. Therefore, we use an MLP network with the same depth and width for the first 30 dimensionless time units, training for the maximum $1000$ epochs per snapshot to obtain a coarse but inexpensive prediction, and switch to the Fourier-feature network with 5000 epochs per snapshot. An early-stopping criterion is also imposed on the training: once all constraints in formulation~\eqref{eq:steady_flow_cylinder_formulation} fall below $10^{-6}$, the training for the current time instant is terminated. In addition, transfer learning is adopted to exploit the similarity between consecutive snapshots: rather than being initialized randomly, the network trained at the previous time instant serves as the pre-trained model for the next one. 

\begin{figure}[!htb]
    \centering
    \includegraphics[width=1.\linewidth]{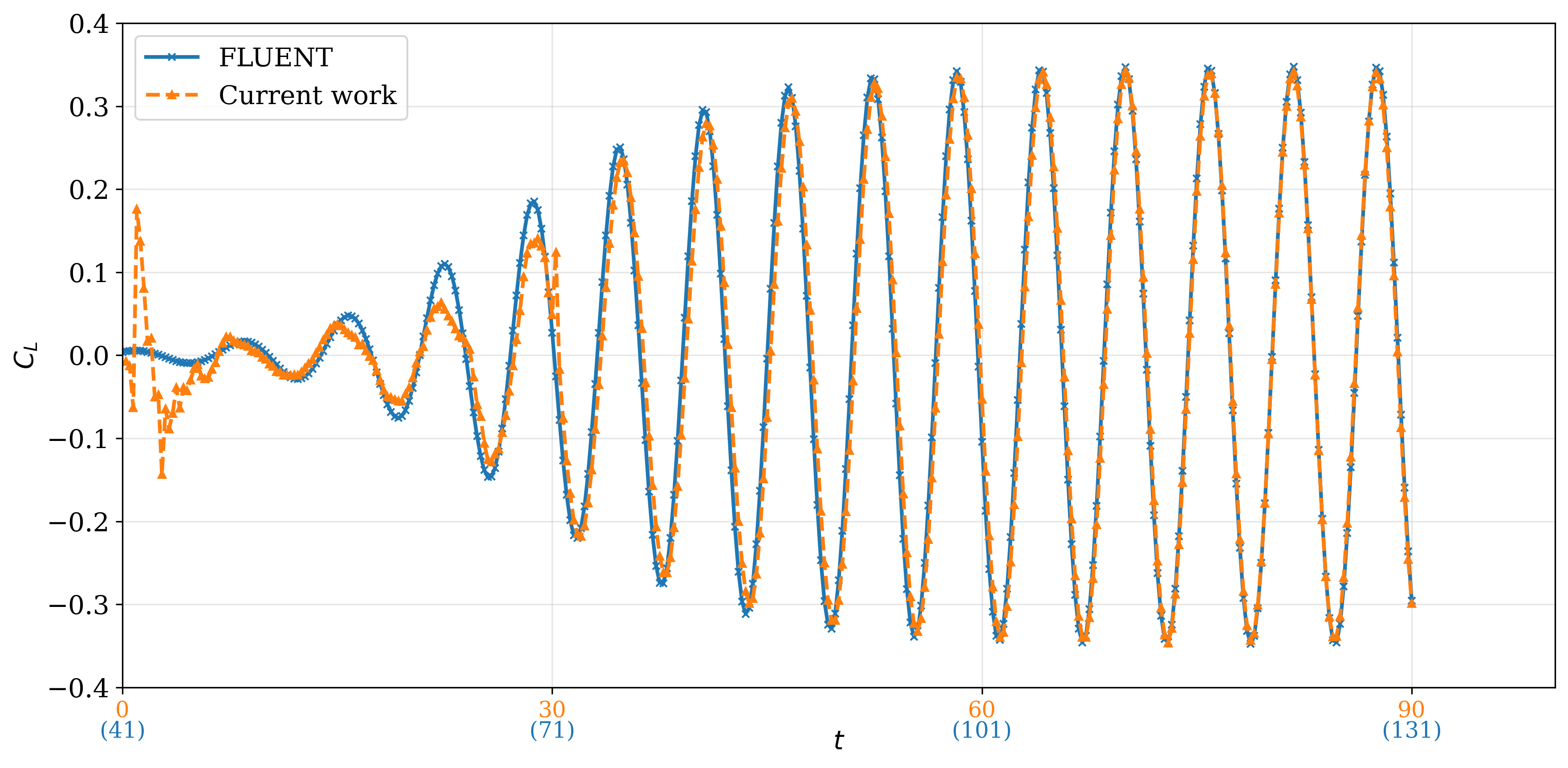}
    \caption{Unsteady flow over a cylinder at $Re=100$: temporal evolution of the lift coefficient $C_L$  from the pressure-Poisson-based formulation and Ansys Fluent, using identical geometry and time step size. The Fluent time axis is shifted to align the periodic regimes for direct comparison. Time axis is normalized by $t_{\infty} =D/U_{in}$.}
    \label{fig:unsteady_cylinder_cl_time_comp}
\end{figure}

The shedding frequency is sensitive to domain size and boundary condition type \citep{Williamson1996}. For a fair comparison, we performed Ansys Fluent simulations using the same time step and domain size. At the outlet, we choose an outflow boundary condition. A mesh refinement study was conducted, and the finest-mesh solution is used for comparison. This mesh is automatically generated with a quadrilateral-dominant method and a base element size of $0.15$; an inflation layer around the cylinder, with a first-cell height of $0.025$, a growth rate of $1.2$, and $30$ layers, yields a total of approximately $2 \times 10^{4}$ cells. The solver uses the pressure-based coupled algorithm with second-order spatial discretization, $100$ inner iterations per time step, and a convergence criterion of $10^{-6}$.

Figure~\ref{fig:unsteady_cylinder_cl_time_comp} compares the temporal evolution of $C_L$ from the proposed pressure-Poisson-based formulation against Ansys Fluent. Since the two approaches are fundamentally different, the onset of shedding from a zero initial field cannot be controlled. To align the periodic regimes for direct comparison, the Ansys Fluent time axis is shifted by
$41$ dimensionless time units, where time is normalized by $t_{\infty} =D/U_{in}$  Once periodic shedding is fully established at $t \approx 60t_{\infty}$, we can observe that the two curves are in excellent agreement. 

Because the solution is obtained via stochastic optimization, the peak value of the lift coefficient $C_L$ varies slightly across shedding cycles; we therefore average it over the last five periods. The mean value of $0.3408 \pm 0.0013$ is in close agreement with the Ansys Fluent prediction of $0.3458 \pm 0.0016$.  Both methods yield a shedding period of $T = 5.75t_{\infty}$, corresponding to a Strouhal number $St = 1/T \approx 0.174$, consistent with $St=0.1754$ reported in \citet{Calderer2010}. The slight discrepancy is attributable to the finer time step used in that study.

\begin{figure}[!htb]
    \centering
    \subfloat[]{\includegraphics[width=1.\textwidth]{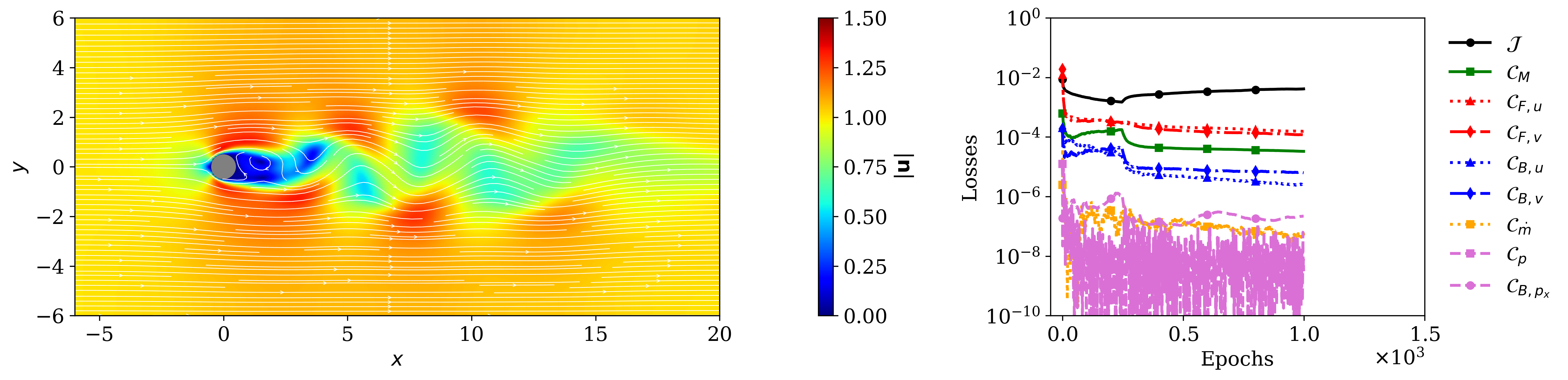}} 
    \\
    \subfloat[]{\includegraphics[width=1.\linewidth]{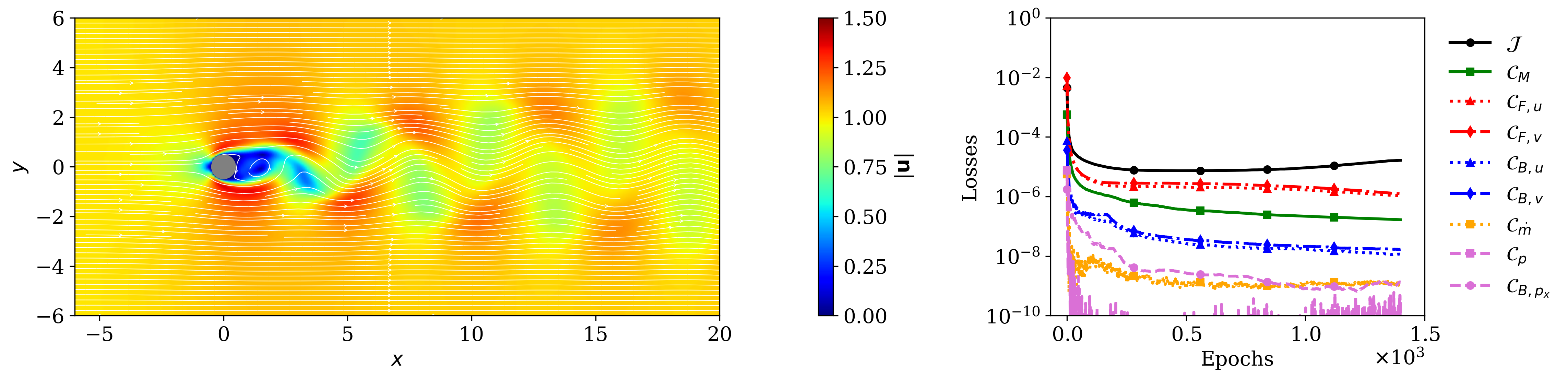}}
    \caption{Unsteady flow over a cylinder at $Re=100$: distribution of predicted velocity magnitude with streamline (left) and the corresponding training history of the loss terms (right): (a) Transient stage at $t=30t_{\infty}$, obtained with the MLP network; (b) periodic stage at $t=80t_{\infty}$, obtained with the Fourier-feature network.}
    \label{fig:unsteady_cylinder_pred_loss_evol}
\end{figure}

Next, we assess the performance of the MLP and Fourier-feature networks during  transient and the periodic stages of the flow evolution. Figure~\ref{fig:unsteady_cylinder_pred_loss_evol} shows the contours of the predicted velocity magnitude, the pressure field and the corresponding loss history at $t=30t_{\infty}$ and $t=80t_{\infty}$. In panel (a), the prediction at $t=30t_{\infty}$ already exhibits the early stage of vortex shedding, despite the relatively high loss values after $1000$ epochs: the objective remains around $10^{-2}$ and the momentum constraints barely fall below $10^{-4}$.
Nevertheless, this coarse prediction provides a good initialization for the subsequent fine-tuning stage with the Fourier-feature network.
This behavior also suggests that the stochastic optimization error of the MLP network triggers the onset of shedding earlier than the discretization error of the finite-volume method in Fluent, as reflected in the shifted time axis of Figure~\ref{fig:unsteady_cylinder_cl_time_comp}.

The fully developed periodic regime is shown at $t = 80t_{\infty}$ in Fig.~\ref{fig:unsteady_cylinder_pred_loss_evol}(b), where a well-defined vortex street extends throughout the wake with no visible influence from the outlet boundary condition. The loss terms converge in roughly $1500$ epochs, with the objective falling below $10^{-4}$ and all constraints satisfying the early-stopping criterion of $10^-6$.

\begin{figure}[!htb]
    \centering
    \subfloat[]{\includegraphics[width=0.45\textwidth]{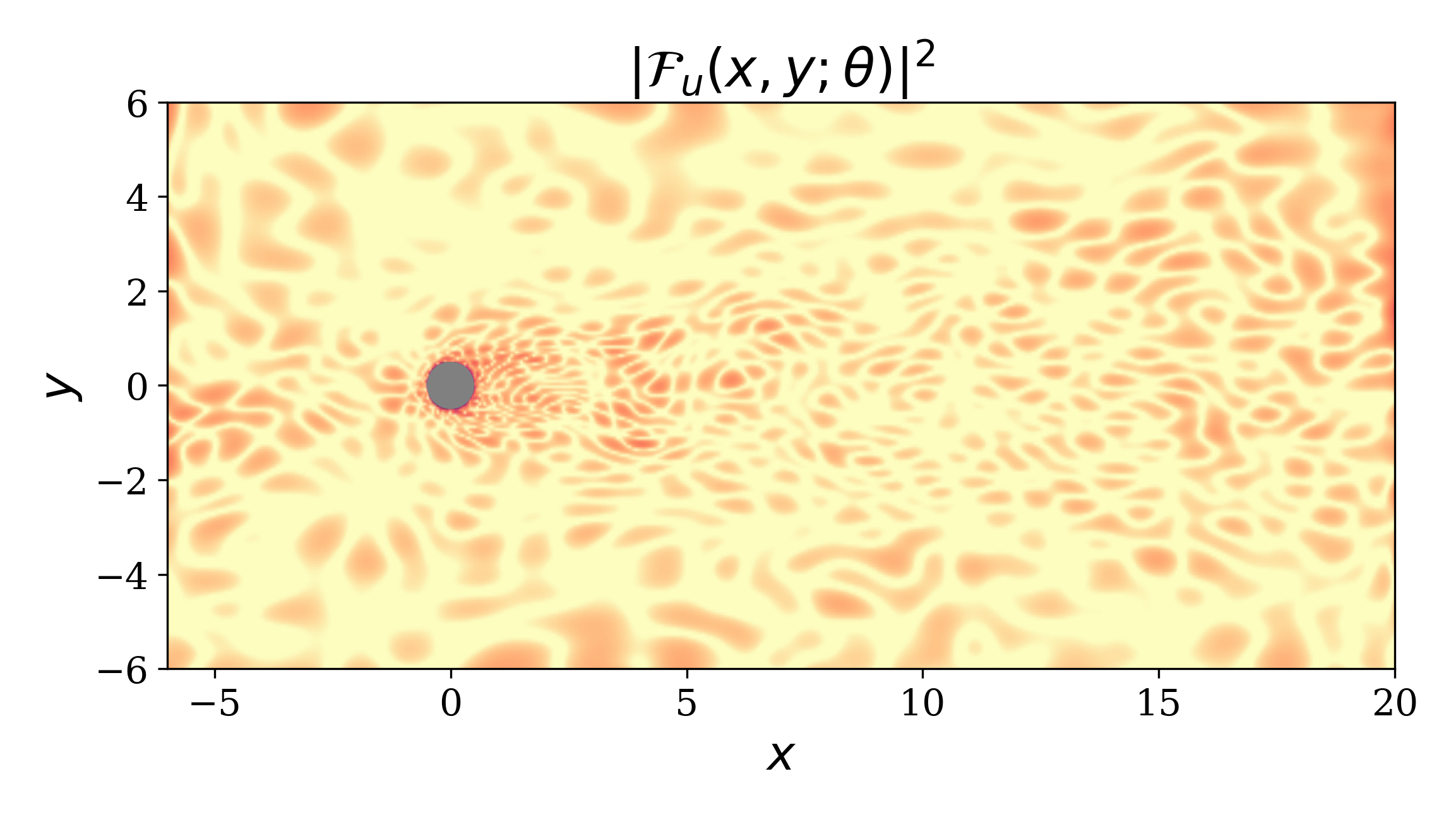}} 
    \quad
    \subfloat[]{\includegraphics[width=0.45\linewidth]{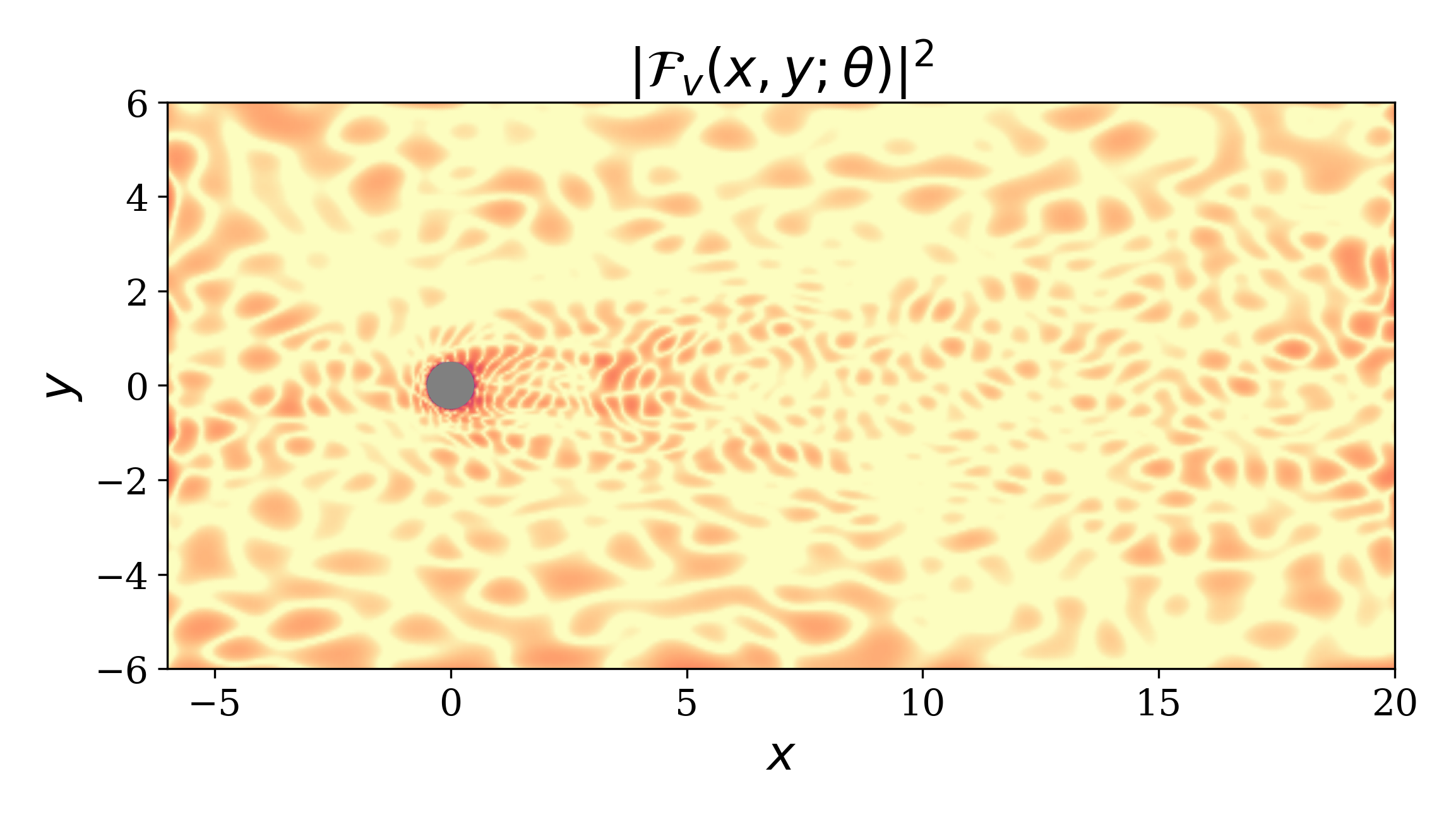}}
    \\
    \subfloat[]{\includegraphics[width=0.53\linewidth]{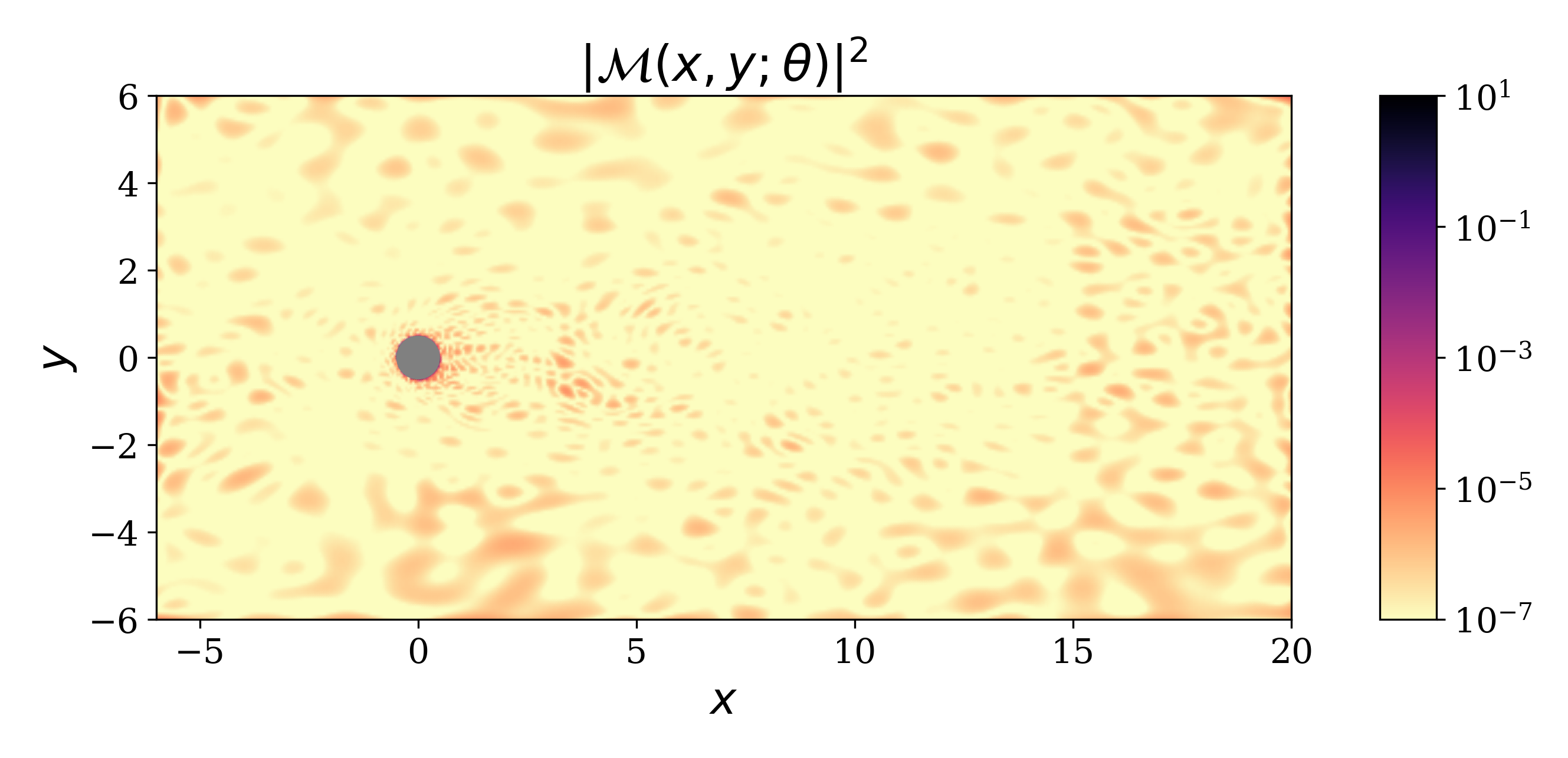}}
    \caption{Unsteady flow over a cylinder at $Re=100$: unseen distribution of squared residuals of (a) the $u$-momentum (b) the $v$-momentum and (c) the mass conservation equations at $t=80 t_\infty$.}
    \label{fig:unsteady_cylinder_unseen_eval}
\end{figure}

Figure~\ref{fig:unsteady_cylinder_unseen_eval} presents the unseen distribution of squared residuals of the $u-$, $v-$ momentum equations and the divergence of the velocity field at $t=80t_{\infty}$, using $500\times200$ mesh points. Note that the unseen distribution does not include the training points. Owing to the limited number of collocation points and early-stopping criterion for constraint satisfaction — necessary to keep time-dependent simulation costs manageable — regions of elevated error in the momentum residuals, in panels~(a) and~(b), are noticeably larger than those in the steady flow case shown earlier in Fig.~\ref{fig:flow_cylinder_re40_unseen_sqrd_res_comp}(c). The divergence field in Fig.~\ref{fig:unsteady_cylinder_unseen_eval}(c) is of order $10^{-7}$ throughout most of the domain, particularly in the wake, demonstrating that the proposed method enforces mass conservation to strict tolerances.

\section{Conclusion}\label{sec:Conclusion}
In this study, we presented a neural incompressible flow solver for steady and unsteady canonical flow problems that requires no labeled data or supervised pretraining. Building on the constrained optimization formulation of PECANN \citep{PECANN_2022}, we introduced an objective function in which the pressure Poisson residual is minimized subject to the momentum equations, the divergence-free condition, and boundary conditions imposed as equality constraints. Enforcing these heterogeneous constraints to the strict tolerances required for physically consistent predictions relied on the conditionally adaptive augmented Lagrangian method (CA-ALM) \citep{hu2026caalm}. Together, these ingredients enabled predictions up to $Re=7,500$ for the lid-driven cavity, and, starting from a randomly initialized network and without external perturbations, the spontaneous onset of periodic vortex shedding past a circular cylinder.

In the proposed pressure-Poisson-based formulation, a single network parameterized both the velocity and pressure fields of the unsteady three-dimensional Navier--Stokes equations. Two further ingredients proved necessary at high Reynolds number. First, an adaptive vanishing entropy viscosity, which we propose here, stabilized the early stages of training without affecting the final predictions. Second, a single Fourier feature mapping of the input coordinates supplied the expressiveness required for high-Reynolds-number and unsteady flows; without it, training failed to converge. We considered both spatio-temporal and discrete-time formulations, and found the discrete-time formulation better suited to capturing long-time flow evolution.

Flow problems with inlet and outlet boundaries required careful treatment of the outflow condition, whereas the velocity conditions used in conventional CFD methods were sufficient at the inlet, symmetry, and wall boundaries. An ablation study over combinations of Neumann conditions on pressure and velocity showed that enforcing global mass flow rate conservation at the outlet was necessary in every case tested, in line with the practice adopted in conventional incompressible flow solvers. The best agreement with benchmark data came from a zero-Neumann pressure condition in the outlet-normal direction combined with a global mass-flow-rate constraint and no additional conditions on velocity; this treatment was used for all steady and unsteady simulations. In all cases, the pressure was anchored at an arbitrary point. Since pressure enters the momentum equations only through its gradient, its absolute level is arbitrary; left unanchored, it can potentially drift to large magnitudes degrading the precision of the computed pressure gradients.

The lid-driven cavity problem has long served as an effective test for CFD methods, particularly in advection-dominated regimes at $Re \gtrsim 1,000$. Extensive validation on 2D lid-driven cavity flow up to $Re=7,500$ showed excellent agreement with reference data across all cases. A baseline formulation that minimized the momentum equation residuals subject to boundary conditions and a divergence-free velocity constraint --- within the same PECANN/CA-ALM framework --- proved ineffective even when equipped with adaptive vanishing entropy viscosity and Fourier feature mapping, underscoring the critical role of the pressure-Poisson-based objective. Further validation on two-dimensional steady backward facing step, three-dimensional unsteady Beltrami flow and both steady and unsteady two-dimensional cylinder flow confirmed the robustness and generality of the proposed method across a range of flow regimes. Most remarkably, the method captured the spontaneous onset of periodic vortex shedding in unsteady cylinder flow without external perturbations, starting from a randomly initialized network.

\textcolor{black}{The proposed neural incompressible flow solver constitutes a fully meshless framework in which all flow variables are represented by neural networks trained on randomly distributed collocation points. While conventional CFD methods remain significantly more efficient for fluid-flow simulations as a forward problem, the present validation establishes a foundation for extending the framework to inverse, parameter-identification, control, and data-driven problems, where neural-network-based approaches can offer distinct advantages through the seamless integration of physical models and observational data.}

\textcolor{black}{For the benchmark problems studied here, the training cost of the neural solver remains approximately one order of magnitude higher than that of conventional commercial CFD solvers. A JAX-based implementation reduced this cost by roughly a factor of three to four (see \ref{sec:app_c}), and further reductions in computational turnaround time may be achieved through domain-decomposition strategies tailored for neural networks \citep{hu_non-overlapping_2025}.} 

\textcolor{black}{Although the solver has been validated on several benchmark problems, including advection-dominated regimes and time-dependent three-dimensional flows, its capacity to simulate three-dimensional turbulent flows remains to be assessed. The neural incompressible flow solver is available as an open-source project at} \url{https://github.com/HiPerSimLab/PECANN-NIS3D}.

\section*{Acknowledgments}
This material is based upon work supported by the National Science Foundation under Grant No. 1953204 and in part by University of Pittsburgh Center for Research Computing and Data, RRID:SCR\_022735, through the resources provided. Specifically, this work used the H2P cluster, which is supported by NSF Award No. OAC-2117681.

%%%%

\appendix
 
\section{Steady lid-driven cavity at $Re=5,000$}\label{sec:app_a}
\begin{figure}[!htb]
    \centering
    \includegraphics[width=0.9\linewidth]{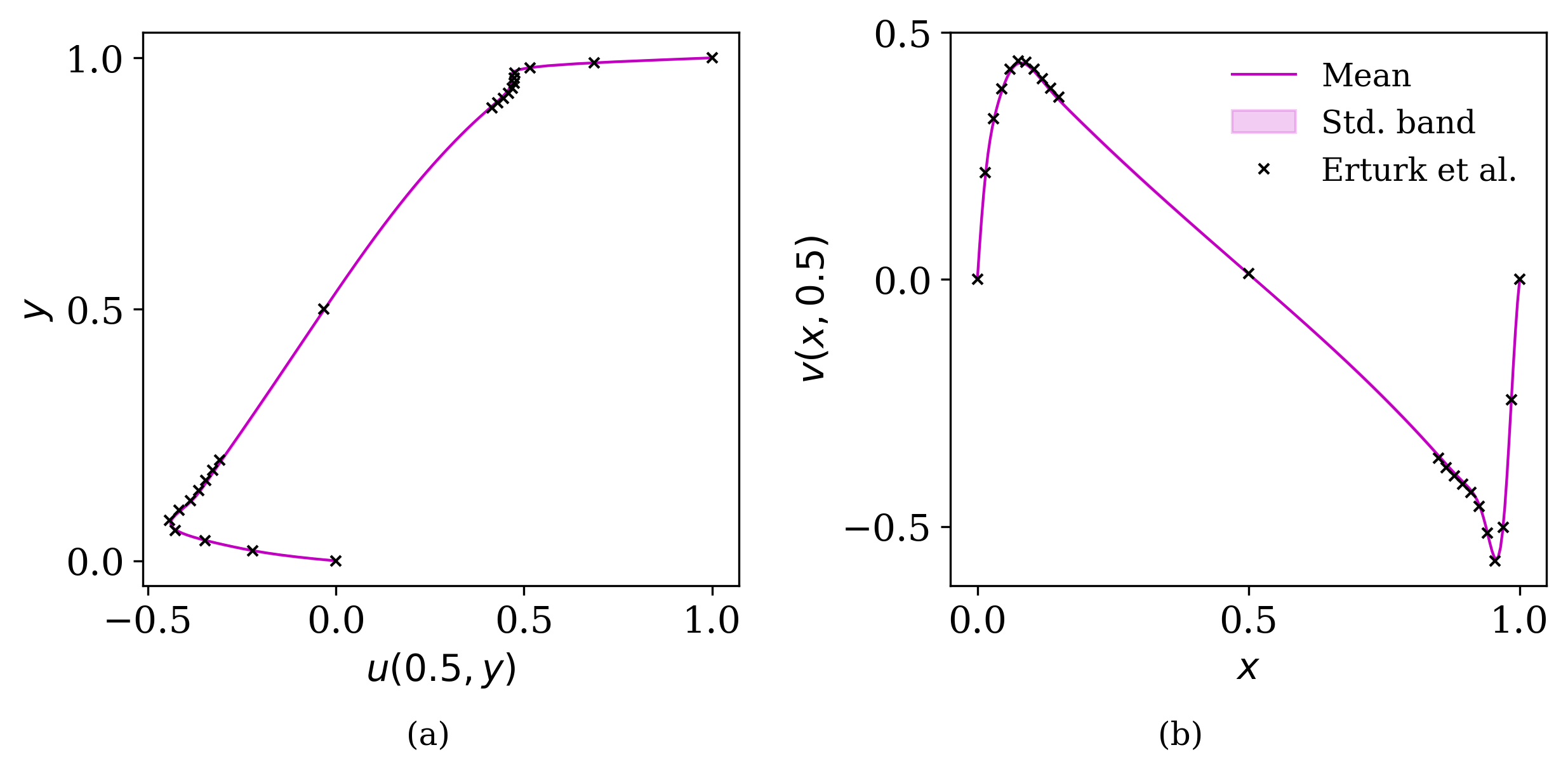}
    \caption{Lid-driven cavity flow at $Re = 5000$: mean and standard deviation of predicted (a) horizontal velocity $u$ along $x=0.5$ and (b) vertical velocity $v$ along $y=0.5$, compared to the benchmark data of \citet{erturk_numerical_2005}.}
    \label{fig:lid_cavity_re5000_profiles}
\end{figure}

Based on the $Re=2,500$ results in Section~\ref{sec:lid_driven_cavity_flow}, we simulate 
$Re=5,000$ using the pressure-Poisson-based formulation~\eqref{eq:proposed_constrained_flow_problem} with Fourier feature mapping~\eqref{eq:modified_1st_hidden} and adaptive vanishing entropy viscosity $\nu_{a}^{a}$ (Eq.~\ref{eq:adapt_av_update}), retaining the same configuration and collocation points as the $Re=2,500$ case. Figure~\ref{fig:lid_cavity_re5000_profiles} shows excellent agreement with the reference data of \citet{erturk_numerical_2005} for both
$u-$ and $v-$ velocity components.

\begin{figure}[!htb]
    \centering
    \includegraphics[width=0.95\linewidth]{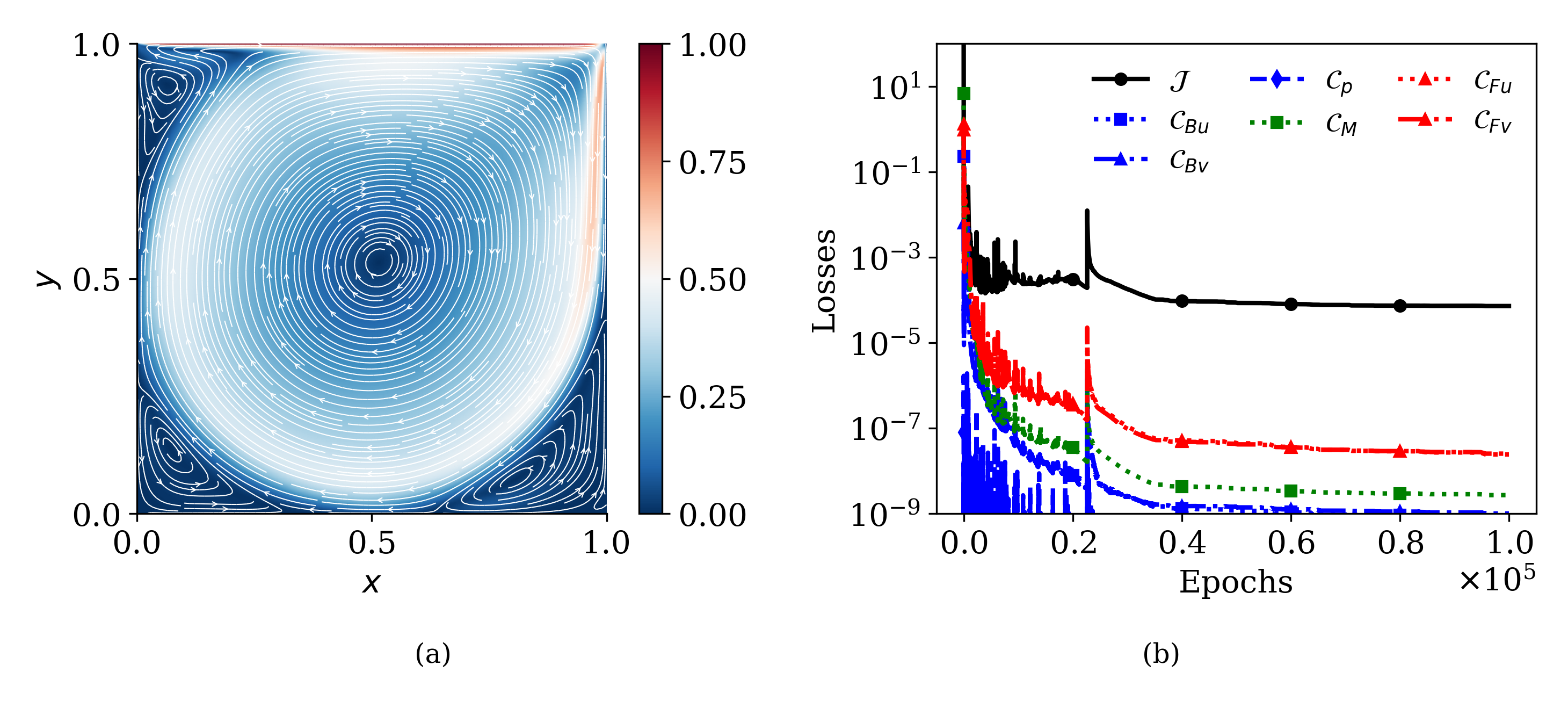}
    \caption{Lid-driven cavity flow at $Re = 5000$: (a) velocity field with streamlines from one Fourier-feature network trial with adaptive vanishing viscosity; (b) corresponding loss evolution.}
    \label{fig:lid_cavity_re5000_contour_loss_evol}
\end{figure}

Figure~\ref{fig:lid_cavity_re5000_contour_loss_evol}(a) shows the contours of the predicted velocity magnitude and streamlines from one trial. At $Re=5000$, the prediction accurately captures the larger vortex near the top-left corner, consistent with the expected strengthening of the circulation compared to the $Re=2500$ case. However, the smallest secondary vortex 
near the bottom-right corner is less distinct, in contrast to the presentation of 
\citet{erturk_numerical_2005}.
Fig.~\ref{fig:lid_cavity_re5000_contour_loss_evol}(b) presents the corresponding evolution of the 
objective and constraint terms. The final spike in the loss curve occurs when the adaptive viscosity 
$\nu_{a}^{a}$ vanishes at approximately $2\times10^{4}$ epochs, after which the optimization 
finally stabilizes around $4\times10^{4}$ epochs.

\section{Steady lid-driven cavity at $Re=7,500$}\label{sec:app_b}
\begin{figure}[!htb]
    \centering
    \includegraphics[width=0.9\linewidth]{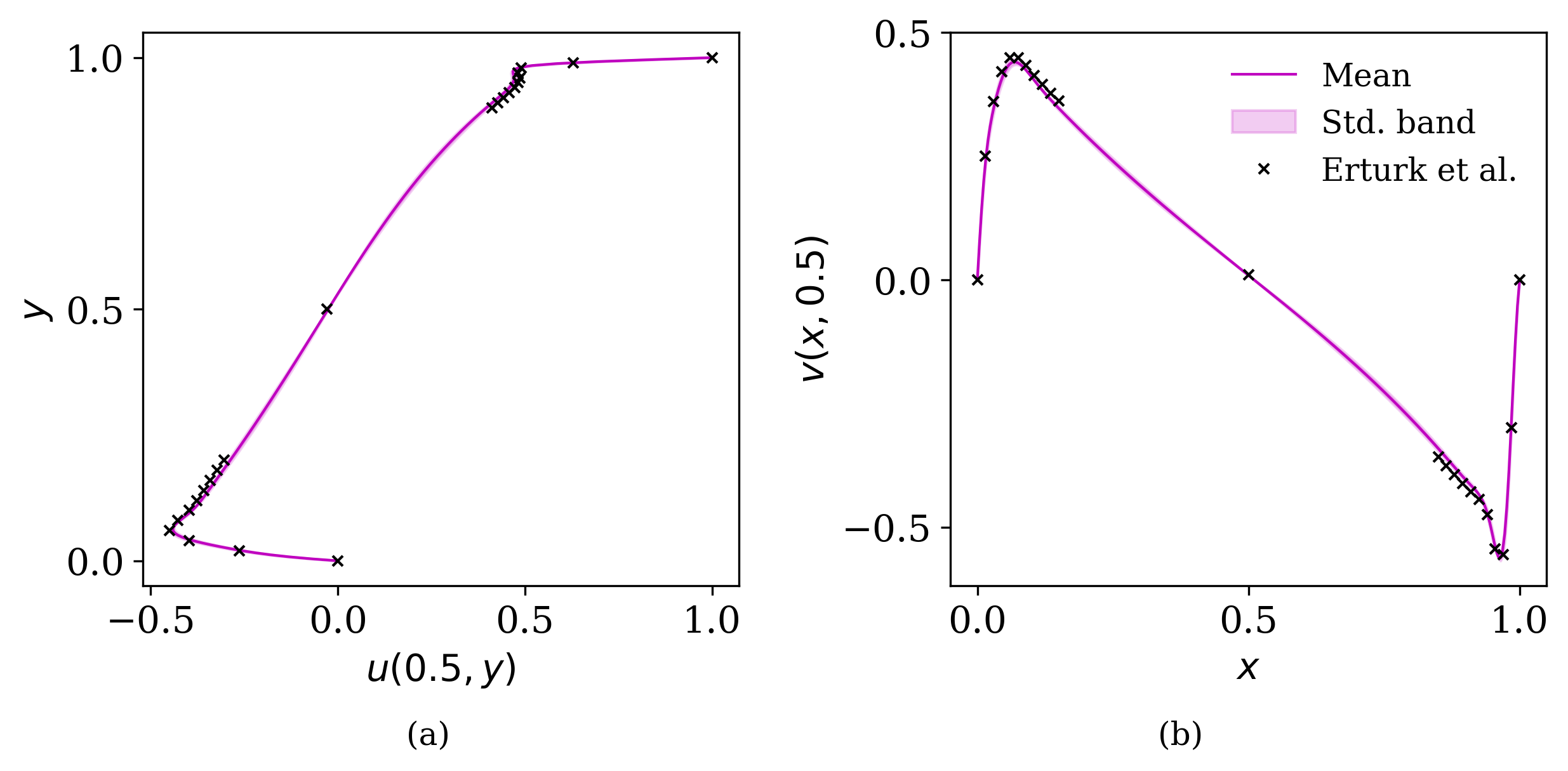}
    \caption{Lid-driven cavity flow at $Re = 7500$. Same legend as in Fig. \ref{fig:lid_cavity_re5000_profiles}.}
    \label{fig:lid_cavity_re7500_profiles}
\end{figure}

The $Re=7,500$ case also reuses the $Re=2,500$ configuration, which already has a high computational cost. Figure \ref{fig:lid_cavity_re7500_profiles}(a) shows the mean $u-$velocity profiles, which are excellent agreement with the reference data. The mean $v-$velocity profile in Fig. \ref{fig:lid_cavity_re7500_profiles}(b) shows a slight mismatch with the reference data. As noted in Section \ref{sec:lid_driven_cavity_flow}, 2D LDC flow exhibits a Hopf bifurcation at $Re\approx8,000$ \citep{auteri2002_LDC_bifurcation, BRUNEAU2006_LDC}. We attribute this minor discrepancy to the proximity of the $Re = 7,500$ to this critical threshold and the reuse of the same configuration from $Re=2,500$ case, which may become insufficient for $Re=7,500$.

\begin{figure}[!htb]
    \centering
    \includegraphics[width=0.95\linewidth]{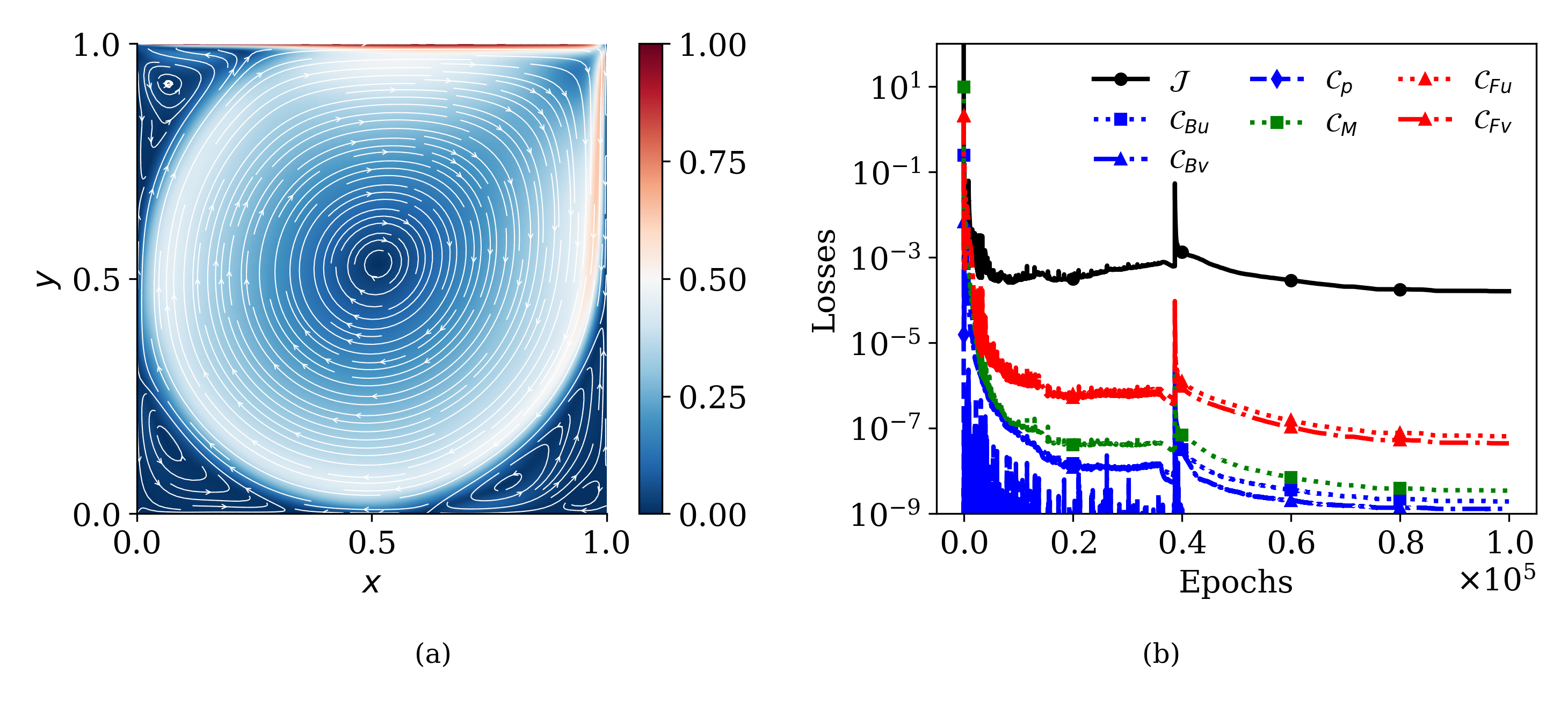}
    \caption{Lid-driven cavity flow at $Re = 7500$. Same legend as in Fig. \ref{fig:lid_cavity_re5000_contour_loss_evol}}
    \label{fig:lid_cavity_re7500_contour_loss_evol}
\end{figure}

The enlarged secondary vortex in the bottom-right corner, shown in Fig.~\ref{fig:lid_cavity_re7500_contour_loss_evol}(a), indicates that the model captures the expected structure of the flow. However, the convergence becomes noticeably slower: the adaptive entropy viscosity vanishes around $4\times10^{4}$ epochs, and the subsequent convergence does not occur until approximately $10^{5}$ epochs.

%%
%\clearpage

\section{\textcolor{black}{Scaling study between PyTorch and JAX
%steady lid-driven cavity at $Re=2,500$
}}\label{sec:app_c}
\begin{figure}
    \centering
    \includegraphics[width=1.\linewidth]{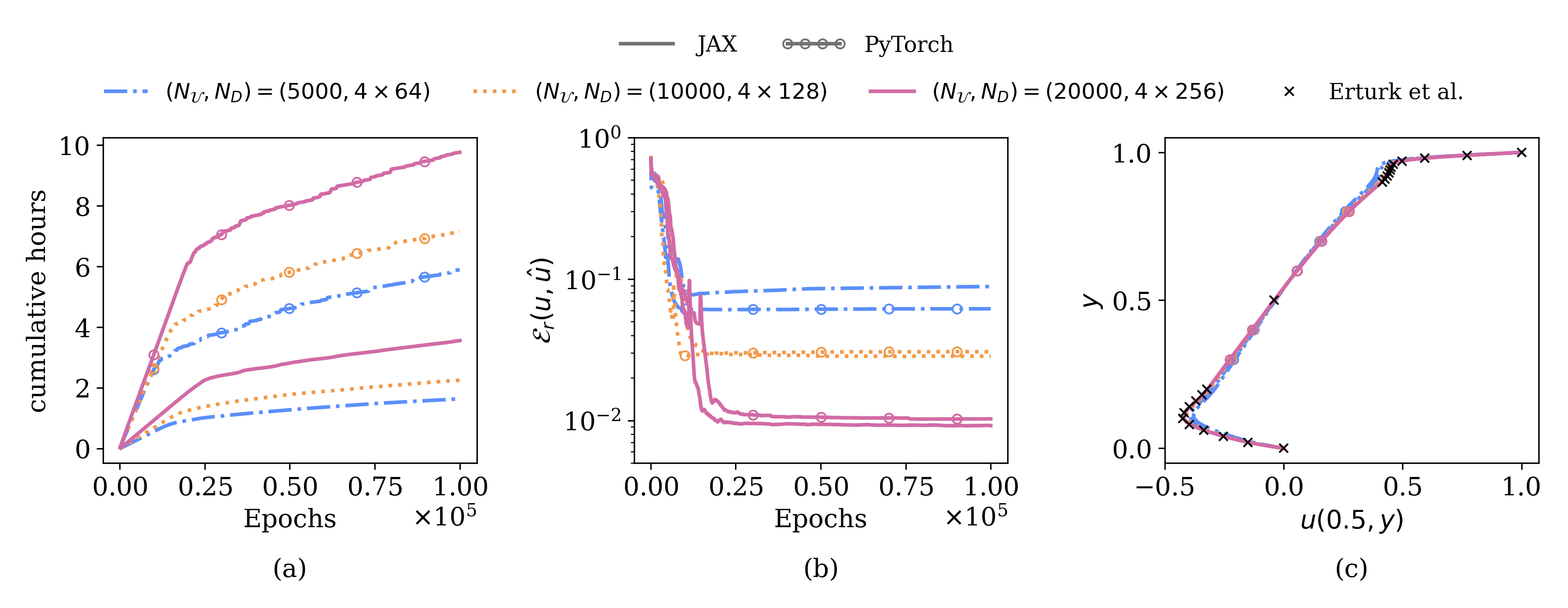}
    \caption{\textcolor{black}{Lid-driven cavity flow at $Re=2500$: scaling with the point distribution in two package implementations. (a) Cumulative wall-clock time against training epochs; (b) relative $\ell^2$ error of the horizontal velocity along the vertical centerline, $\mathcal{E}_r(u,\hat{u})$, calculated with the reference data of \citet{erturk_numerical_2005}; (c) centerline profile $u(0.5,y)$ at the final epochs. Curves with open circles are PyTorch and those without are JAX. The baseline uses $N_{\mathcal{U}}=20{,}000$ collocation points and $N_D=4\times256$ boundary points per edge; the other two halve and quarter both point sets.}}
    \label{fig:lid_cavity_re2500_scaling}
\end{figure}

\textcolor{black}{
To quantify the computational cost of the proposed method, we conduct a scaling study on the steady lid-driven cavity flow at $Re=2500$, not only in the PyTorch code used throughout this work, but also in an equivalent JAX \citep{jax2018github} implementation.
The JAX provides compilation of the entire training step---evaluation of the physics residuals, including the nested reverse-mode differentiation required for the second-order derivatives, the loss and its gradient, and the optimizer update---ahead of time with Accelerated Linear Algebra.
Starting from the configuration used in Sec.~\ref{sec:lid_driven_cavity_flow} ($N_{\mathcal{U}}=20{,}000$ collocation points and $N_D=4\times256$ boundary points), we halve and quarter both point sets, giving $(N_{\mathcal{U}},N_D)=(10{,}000,\,4\times128)$ and $(5{,}000,\,4\times64)$.
All remaining settings, including the L-BFGS optimizer with strong Wolfe line search (based on PyTorch), are identical between the two package frameworks, and every run is performed on a single NVIDIA A100 GPU node.
Figure~\ref{fig:lid_cavity_re2500_scaling} reports, for each point distribution and each package framework, the cumulative wall-clock time, the relative $\ell^2$ error of the horizontal velocity along the vertical centerline, $\mathcal{E}_r(u,\hat{u})$, calculated with the reference data of \citet{erturk_numerical_2005} and the converged centerline profile.
}

\textcolor{black}{
Two regimes are apparent in the timing curves of Fig.~\ref{fig:lid_cavity_re2500_scaling}(a), in both frameworks.
Initially the cost grows linearly, indicating that L-BFGS exhausts its default maximum number of inner iterations at every epoch (based on PyTorch); the cost per epoch is therefore constant and is set by this fixed iteration budget. 
Within each package framework, the three slopes differ only mildly: the finest distribution is the most expensive, while the halved and quartered configurations are nearly indistinguishable, suggesting that at or below $10^{4}$ points the cost is dominated by the number of inner iterations rather than by the number of points. 
Each curve then bends and grows more slowly, because once the loss stagnates the line search terminates after fewer inner iterations, so fewer function and gradient evaluations are required per epoch.
}

\textcolor{black}{
The compiled implementation of JAX reduces the cost substantially and uniformly. After $10^{5}$ epochs it runs require approximately $1.6$, $2.3$ and $3.5$ hours for the quartered, halved and full distributions in Fig.~\ref{fig:lid_cavity_re2500_scaling}(a), with a speed-up of between three and four compared to PyTorch.
The reduction confirms that a large part of the per-epoch cost in PyTorch is interpreter and kernel-launch overhead rather than arithmetic, since it is removed by ahead-of-time compilation of the training step without any change to the algorithm.
The sub-linear scaling with the number of points is preserved: in both frameworks a fourfold increase in the point count raises the total cost by less than a factor of two.
}

\textcolor{black}{
Fig.~\ref{fig:lid_cavity_re2500_scaling}(b) shows that this bend coincides, for each configuration, with the epoch at which $\mathcal{E}_r$ reaches its plateau: near $1\times10^{4}$ epochs for the quartered and halved distributions and near $2\times10^{4}$ epochs for the full one. 
The change in slope of the timing curve is therefore a direct signature of convergence.
Beyond that point the error is essentially flat, so the remaining epochs contribute little accuracy: in PyTorch the useful cost of each run is approximately $3.0$, $4.0$ and $6.5$ hours, against totals of $6.0$, $7.0$ and $10.0$ hours at $10^{5}$ epochs. 
An appropriate stopping criterion would thus roughly halve the reported cost without loss of accuracy.
}

\textcolor{black}{
The converged profiles in Fig.~\ref{fig:lid_cavity_re2500_scaling}(c) are visually indistinguishable from one another, between frameworks, and from the reference data over most of the centerline; only the coarsest distribution departs noticeably in the near-wall region. 
The halved configuration $(10000, 4\times128)$ already reproduces the profile accurately. 
Combined with the early convergence identified above and the speed-up of the compiled implementation, this configuration requires around 1.5 hours to reach its plateau at less than $2.5\times10^{4}$ epochs.
}

\textcolor{black}{
The cost of the present method nonetheless remains above that of a conventional solver, although by less than an order of magnitude: the same problem computed with Ansys Fluent on a structured mesh $256\times256$ achieves a comparable centerline profile in approximately $0.3$~h, on a single core of an AMD EPYC 74F3.
The purpose of this study is therefore not to compete with conventional numerical solvers on forward problems for which they are already efficient, but to characterize how the cost of the proposed method scales with the point distribution and to establish the price of a mesh-free, differentiable formulation whose advantages lie elsewhere: in the assimilation of labeled data, in inverse problems, and in parametric sweeps.
}

\section{\textcolor{black}{Ablation study of adaptive entropy viscosity in steady lid-driven cavity at $Re=2,500$}}\label{sec:app_d}
\begin{figure}
    \centering
    \includegraphics[width=1.\linewidth]{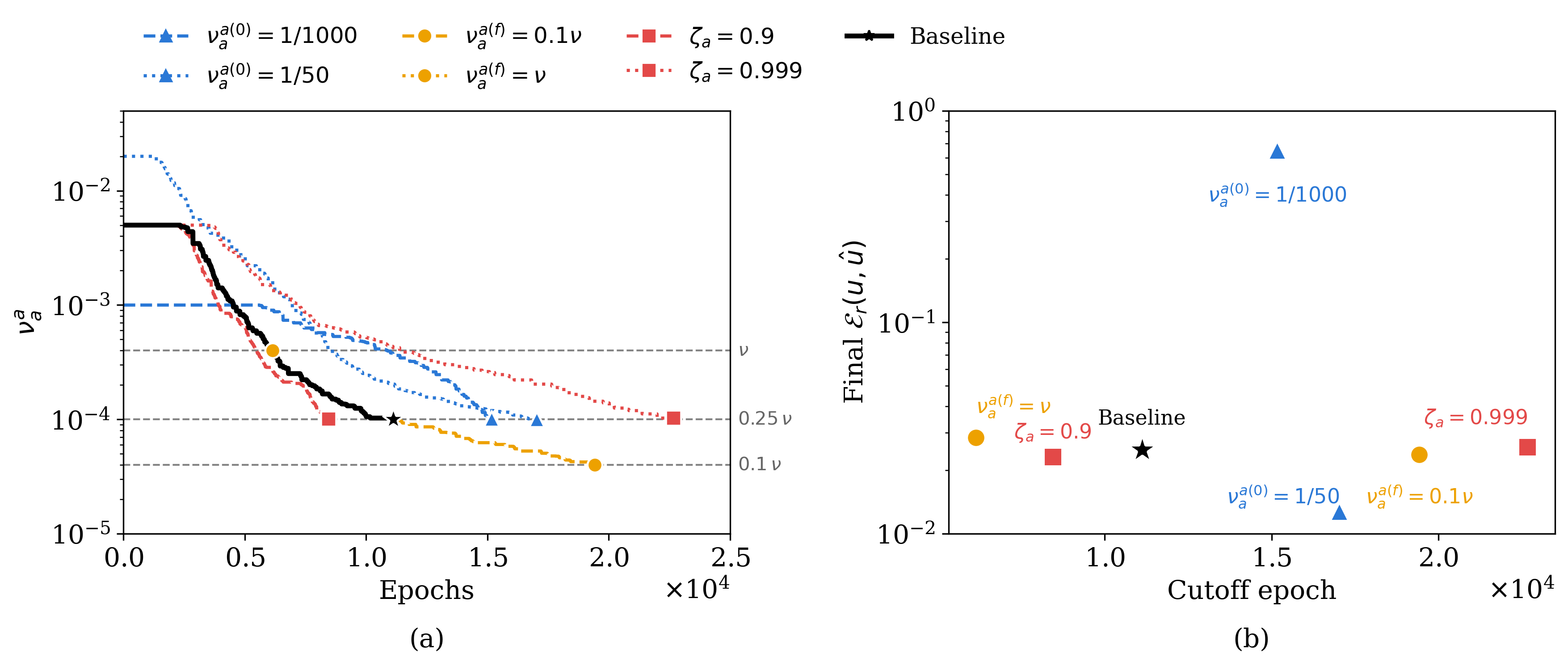}
    \caption{\textcolor{black}{Lid-driven cavity flow at $Re=2500$: (a) Decay of the artificial viscosity $\nu_a^a$ with training epochs; (b) Relative $\ell^2$ error of the horizontal velocity along the vertical centerline after $10^{5}$ epochs, against the cutoff epoch identified in (a), compared to the reference data of \citet{erturk_numerical_2005}. Each curve varies one hyperparameter about the baseline $\nu_a^{a(0)}=\tfrac{1}{200}$, $\zeta_a=0.99$, $\nu_a^{a(f)}=\tfrac{\nu}{4}$, with all other settings unchanged.}}
    \label{fig:lid_cavity_re2500_ablation}
\end{figure}

\textcolor{black}{
The formulation of adaptive vanishing viscosity in Eq.~\eqref{eq:adapt_av_update} introduces three hyperparameters: the initial viscosity $\nu_a^{a(0)}$, the smoothing coefficient $\zeta_a$ and the cutoff viscosity $\nu_a^{a(f)}$. 
To assess the sensitivity of the method to them, we repeat the steady lid-driven cavity at $Re=2500$ while varying each in turn about the baseline values $\nu_a^{a(0)}=\tfrac{1}{200}$, $\zeta_a=0.99$ and $\nu_a^{a(f)}=\tfrac{\nu}{4}$, holding all remaining settings fixed.
Figure~\ref{fig:lid_cavity_re2500_ablation}(a) shows the resulting decay of $\nu_a$, and Fig.~\ref{fig:lid_cavity_re2500_ablation}(b) plots the final relative $\ell^2$ error of the horizontal velocity against the epoch at which the artificial viscosity is switched off.
}

\textcolor{black}{
The three hyperparameters act on the decay history in distinct and predictable ways. 
Increasing $\zeta_a$ lengthens the memory of the moving average and therefore delays the cutoff, from approximately $0.9\times10^{4}$ epochs when $\zeta_a=0.9$ to $2.3\times10^{4}$ when $\zeta_a=0.999$. 
Lowering $\nu_a^{a(f)}$ requires $\nu_a^a$ to fall further before it is retired: at $\nu_a^{a(f)}=\nu$ the artificial viscosity is removed as soon as it reaches the physical value $\nu = 1/Re$, after roughly $0.6\times10^{4}$ epochs, whereas at $\nu_a^{a(f)}=0.1\nu$ it persists to a tenth of $\nu$ and to almost $2\times10^{4}$ epochs. 
The initial value $\nu_a^{a(0)}$ shifts the starting point of the decay.
}

\textcolor{black}{
Despite a spread of a factor of four in the cutoff epoch, the final accuracy is largely insensitive to these variations. 
Every configuration except one attains a final error between $1.5\times10^{-2}$ and $3.0\times10^{-2}$, bracketing the baseline value of $2.2\times10^{-2}$. 
The exception is $\nu_a^{a(0)}=1/1000$, which fails to converge to the correct solution and remains at an error of order $7\times10^{-1}$. 
At this setting, the initial artificial viscosity corresponds to a Reynolds number of $1000$, so it provides too little momentum diffusion to regularize the early training; the optimizer settles in a spurious local minimum from which the subsequent decay of $\nu_a^a$ cannot recover. 
Conversely, the more diffusive start $\nu_a^{a(0)}=1/50$ doesn't impact the latter accuracy ($1.5\times10^{-2}$).
}

\textcolor{black}{
From these observation, the method is robust to $\zeta_a \in \{0.9, 0.99, 0.999\}$ and $\nu_a^{a(f)} \in [0.1\nu, \nu]$ across the ranges tested, and tolerates an over-estimate of $\nu_a^{a(0)}$, but an under-estimate is fatal. 
We therefore recommend initializing $\nu_a^a$ corresponding to the case of Reynolds number 200, which the baseline $\nu_a^{a(0)}=1/200$, and leaving $\zeta_a=0.99$ as the same value in CA-ALM algorithm, and $\nu_a^{a(f)}=\nu/4$ once the artificial viscosity is relatively smaller than the actual values as the default values.
}

\section*{Declaration of generative AI and AI-assisted technologies in the writing }
During the preparation of this work the author(s) used Microsoft Copilot and Claude in order to assist with improving the clarity and quality of the English language. After using this tool/service, the author(s) reviewed and edited the content as needed and take(s) full responsibility for the content of the publication.

%\bibliographystyle{model1-num-names}
%\bibliography{citations}
\end{document}